\documentclass{ws-ijgmmp}

\begin{document}

\markboth{Zi-Hua Weng}
{Forces in the complex octonion curved space}

%
\catchline{}{}{}{}{}
%

\title{Forces in the complex octonion curved space
}

\author{Zi-Hua Weng
}

\address{School of Physics and Mechanical \& Electrical Engineering
\\
Xiamen University, Xiamen 361005, China
\\
\email{xmuwzh@xmu.edu.cn
}
}



\maketitle

\begin{history}
\received{(Day Month Year)}
\revised{(Day Month Year)}
\end{history}

\begin{abstract}
The paper aims to extend major equations in the electromagnetic and gravitational theories from the flat space into the complex octonion curved space. Maxwell applied simultaneously the quaternion analysis and vector terminology to describe the electromagnetic theory. It inspires subsequent scholars to study the electromagnetic and gravitational theories with the complex quaternions/octonions. Furthermore Einstein was the first to depict the gravitational theory by means of tensor analysis and curved four-space-time. Nowadays some scholars investigate the electromagnetic and gravitational properties making use of the complex quaternion/octonion curved space. From the orthogonality of two complex quaternions, it is possible to define the covariant derivative of the complex quaternion curved space, describing the gravitational properties in the complex quaternion curved space. Further it is possible to define the covariant derivative of the complex octonion curved space by means of the orthogonality of two complex octonions, depicting simultaneously the electromagnetic and gravitational properties in the complex octonion curved space. The result reveals that the connection coefficient and curvature of the complex octonion curved space will exert an influence on the field strength and field source of the electromagnetic and gravitational fields, impacting the linear momentum, angular momentum, torque, energy, and force and so forth.
\end{abstract}


\section{Introduction}

In the electromagnetic theory, Maxwell applied simultaneously the quaternion analysis and vector terminology to describe the electromagnetic properties. Hamilton invented the algebra of quaternions in 1843. During two decades from that time on, the quaternion was separated into the scalar and vector parts. In his studies, J. C. Maxwell mingled the quaternion analysis and vector terminology to depict electromagnetic properties in 1873. Presently some scholars apply the vector terminology to depict the physical properties of electromagnetic and gravitational fields, while some scholars begin to study the electromagnetic theory \cite{morita, edmonds} and gravitational theory \cite{majernik, rawat} with the algebra of quaternions.

Graves and Cayley invented independently the octonions, which can be considered as an ordered couple of quaternions. Some scholars applied the octonion analysis \cite{gogberashvili, demir} to describe the electromagnetic and gravitational theories \cite{mironov, negi}. The study reveals that the octonions (or the standard octonions, rather than the split octonions) can be divided into two parts, the quaternions and $S$-quaternions (short for the second quaternion), and their coordinate values are allowed to be complex numbers. And the $S$-quaternion in an octonion is similar to the imaginary number in the complex number. In the complex octonion flat space, the complex quaternion space is appropriate to describe the gravitational property, meanwhile the complex $S$-quaternion space is proper to depict the electromagnetic property (Appendix A). Further these properties can be extended from the complex octonion flat space \cite{weng1} to the complex octonion curved space. That is, the complex octonion curved space is possible to describe the gravitational and electromagnetic fields. The complex quaternion curved space fits for studying gravitational properties, and the complex $S$-quaternion curved space is apt for researching electromagnetic properties.

\subsection{Existing studies}

After the tensor analysis \cite{bishop, morandi} was evolved from the vector analysis and other theories, Einstein \emph{et al.} adopted the tensor analysis and pseudo-Riemannian space theory to study the gravitational properties in the curved four-space-time. Subsequently the quaternion and octonion curved spaces are applied to investigate physical properties of electromagnetic and gravitational fields in some situations. Marques-Bonham \cite{marques} developed the geometrical properties of a flat tangent space-time local to the manifold of the Einstein-Schr\"{o}dinger nonsymmetric theory on an octonionic curved space. Dundarer \cite{dundarer} defined a four-index antisymmetric and non-Abelian field, which satisfies a self-duality relation in eight-dimensional curved space. Tsagas \cite{tsagas} studied the evolution of electromagnetic fields in curved space-times, and calculated the wave equations. Castro \cite{castro} proposed a nonassociative octonionic ternary gauge field theory, based on a ternary bracket. Demir \cite{demir2} formulated the Maxwell-Proca type field equations of linear gravity in terms of hyperbolic octonions (split octonions). Chanyal \cite{chanyal} \emph{et al.} described the octonion formulation of Abelian/non-Abelian gauge theory in terms of the Zorn vector matrix realization. Kalauni \cite{kalauni} \emph{et al.} obtained the fully symmetric Dirac-Maxwell's equations as one single equation by using the matrix presentation, with the help of the algebraic properties of quaternions and octonions.

In the existing electromagnetic and gravitational theories described with the curved spaces, there are mainly three description methods as follows:

1) Curved four-space-time. Within the General Theory of Relativity in 1915, Einstein introduced the method of pseudo-Riemannian space to study gravitational properties in the curved four-space-time. Later some scholars attempt to apply the pseudo-Riemannian space to explore various properties of gravitational field, and even the electromagnetic field. But this method is not successful in depicting convincingly the physical property of electromagnetic field.

2) Real octonion curved space. Some scholars described the physical properties of the gravitational and electromagnetic fields in the flat space, by means of the real quaternion/octonion flat space. However this method was found to be inadequate finally \cite{weng2}. And it is not suitable to describe the gravitational and electromagnetic fields in the curved space either, making use of the real quaternion/octonion curved space. One reason is that the definition of the arc length for this space deviates from the space-time interval in the physics.

3) Complex octonion curved space. The complex quaternion/octonion flat space is able to depict the physical property of gravitational and electromagnetic fields in the flat space. Some scholars research gravitational and electromagnetic properties in the curved space, depending on the complex quaternion/octonion curved space. On the basis of the locally tangent frame and tensor analysis, the paper can explore the characteristics of complex quaternion/octonion curved space. There are a few contemporary results at present. The study reveals that the definition of the arc length for this space is identical with the space-time interval in the physics. The connection coefficient and curvature of the complex quaternion/octonion curved space, which are similar to the field strength and source, will exert an influence on the object's movement to a certain extent, no matter what kind of reason it is to result in the space curving evidently. On the contrary, measuring the field strength and source relevant to the gravitational and electromagnetic fields, one may be able to ascertain the deviation amplitude between the complex quaternion/octonion curved space and its flat space.

Making use of the comparison, the paper found that the four-dimensional Riemannian space corresponds to the four-dimensional Euclidean space. But both of them are not suitable to be applied to describe directly the physical properties of gravitational or electromagnetic field. The four-dimensional pseudo-Riemannian space corresponds to the four-dimensional pseudo-Euclidean space (or Minkowski space, four-space-time), and is appropriate for describing gravitational properties in the General Theory of Relativity. Similarly the real quaternion/octonion curved space corresponds to the four-dimensional Riemannian space, and is unsuitable to depict directly the physical property of gravitational or electromagnetic field either. In contrast, the complex quaternion/octonion curved space corresponds to the four-dimensional pseudo-Riemannian space, and is fit for researching the gravitational and electromagnetic properties. On the basis of this curved space, the paper will study various properties of gravitational and electromagnetic fields in the complex quaternion/octonion curved space, including the field potential, field strength, field source, linear momentum, angular momentum, torque, and force and so forth.

\subsection{Two topics}

In the curved four-space-time of the General Theory of Relativity, the scholars delved into the gravitational theory \cite{fischbach, adelberger}, attaining many scientific achievements. But there are still a few things to worry about.

1) Astrophysical jets. The scholars have being doubted the validity and relevant deductions of E\"{o}tv\"{o}s experiment for a long time \cite{baessler, turyshev}. They not only inspect the E\"{o}tv\"{o}s experiment in the laboratory again and again \cite{moffat}, but also measure repeatedly the gravitational constant in the astronomical observatory \cite{liu}. One recent theoretical result points out that the E\"{o}tv\"{o}s experiment has never been validated under the strong electromagnetic circumstance, suspecting that the gravitational mass may be varied with fluctuation of electromagnetic strength. Further the variable gravitational mass is able to explain the phenomenon of astrophysical jets \cite{weng3}, which is the defeat for the classical gravitational theory or the General Theory of Relativity. Up to the present, this suspicion is an ongoing process, and the answer has not been concluded yet. Obviously the query on the changeability of gravitational mass will impact further the theoretical basis of the General Theory of Relativity. The appeal of improving the General Theory of Relativity is gradually deepening and expanding. This open opinion provides one approach to develop the General Theory of Relativity.

2) Dark matter. An abundance of astronomical observational results reveal that there are a large amount of undetected matters in the galaxies and galaxy clusters. But the classical gravitational theory and the General Theory of Relativity cannot account for the existence of undetected matters. Obviously these two existing theories both are not perfect enough. Nowadays the scholars mainly develop two kinds of investigations for this problem. On the one hand, some scholars propose to modify the laws of gravity established by I. Newton and A. Einstein (such as MOND), attempting to explain the anomalous observations. But this method cannot account for the properties of galaxy clusters. On the other hand, some scholars introduce the concept of `dark matter', explaining the rotational velocity curves of galaxies and the gravitational lensing of galaxy clusters. All of these endeavors state that the classical gravitational theory and the General Theory of Relativity both may encounter the puzzling impediment, which may be very tough to overcome.

3) Curved space. The human being often breaks into exclamations of wonder of scientific laws, even if we are not able to comprehend why the scientific laws are so intriguing. Fortunately the human being possesses the ability to suppose, measure, and describe the particular contents of scientific laws to a certain extent. In other words, `the reason for the existence of scientific laws' and `the specific contents of scientific laws' are two disparate topics. Similarly the human being may not understand why the space is curved when we are in a curved space, although the scholars are able to measure and describe the bending degree of one space. If we find the reason why a space is flat, we may know why other spaces are curved. Therefore `the reason to curve the space' and `the bending degree of the curved space' are two distinct topics, they should be treated discriminatively. Einstein tried to uncover why the space is curved under certain circumstances, in the General Theory of Relativity. However this paper involves in one distinct topic, and focuses on describing the possible influence of the bending degree of curved space on some equations of the field theory.

The analysis in the above reveals that the concept, viewpoint, and experiment relevant to the curved space are evolving constantly and gradually, along with the development of science and technology. For the General Theory of Relativity, it may be necessary to have a fundamental rethink of existing concepts appropriately, and even supplement a few new results.

1) Distorting the result of the E\"{o}tv\"{o}s experiment. The General Theory of Relativity reckons subjectively that the gravitational mass is unchangeable, and does not contemplate the influence of the energy gradient (that is, one force term) on some other experiments \cite{reasenberg}. However some theories deduce that the gravitational mass may be varied with fluctuation of field strength and so forth. It results in the gravitational mass to depart from the inertial mass with a deviation. The energy gradient relevant to the gravitational or electromagnetic strength will act on the E\"{o}tv\"{o}s experiment. Therefore it is necessary to validate the E\"{o}tv\"{o}s experiment under the strong electromagnetic circumstance.

2) Surmising the reason for curving the space. As a pioneer, Einstein broke fresh ground in the field theory, that is, the curved space. This theory surmises subjectively that the curvature tensor is directly proportional to the energy-momentum tensor, deducing the field equations. However the field equations should be verified continually in the experiment. The field equations are doubtful, and call for more collateral evidences to support its claim. You can even guess that there may be more than one reason to curve the space.

3) Origin of mass. This is one topic that the General Theory of Relativity has never been touched on. At present a few scholars are engaging in some projects to figure out the origin and component of the mass. Undoubtedly this is one underlying challenge for some propositions in the General Theory of Relativity.

In a word, the paper will focus on studying the influence of curved space on the physical quantity, rather than try to find out the reason why the space is curved in the physics, by means of the affine frame of complex octonion curved space.

\section{Orthogonal affine frame}

In this paper, it is necessary to introduce the tangent-frame component, $C_\nu$ , of one point $P$ in the curved space, in order to define the component of a physics quantity in the curved coordinate system. And the tangent frame system $\{ C_\nu \}$ belongs to the affine frame. In general the tangent-frame components are neither equal-length, nor perpendicular to each other. For the pseudo-Riemannian space in the General Theory of Relativity, the tangent-frame component may be a vector or a scalar. In the quaternion or octonion curved space, the tangent-frame component, $C_\nu$ , may be the quaternion or octonion respectively. $\nu = 0, 1, 2, 3, 4, 5, 6, 7$.

In the physics, the most common coordinate systems are orthogonal and even orthonormal, to give prominence to the physical contents and lower down the degree of difficulty of the mathematics in the physics theories. That is, not all affine frames are suitable to be chosen as the curved coordinate system, for the gravitational and electromagnetic fields in the curved space. And it is essential to demarcate and filter out the appropriate affine frames further. When we study the physics quantity and relevant properties in the curved space, it is proper to stipulate to choose the orthogonal affine frame (or the orthogonal curved coordinate system), for the sake of reducing the correlative mathematical difficulty (such as, the metric tensor and invariants). On the other hand, the physics quantity in the orthogonal affine frame can be transformed into one in an orthogonal and equal-length affine frame, in order to draw a comparison between the physics quantity in the flat space (see Ref.[9]) with that in the tangent space of the curved space.

According to this arrangement, the gravitational and electromagnetic theories can be extended from the flat space into the curved space. Firstly, transform the gravitational and electromagnetic theories in the flat space into that under the orthogonal and equal-length affine frame of the tangent space in the curved space; Secondly, transform the gravitational and electromagnetic theories in the orthogonal and equal-length affine frame into that in the orthogonal affine frame; Thirdly, in the orthogonal affine frame, study the influence of the curved space on the force and so forth in the gravitational and electromagnetic fields.

In the curved space, the physics quantity will contain several parameters of the curved space. a) The integrating function, $\mathbb{X}$, of field potential can be chosen as the first-rank tensor, and then the field potential, $\mathbb{A}$, as the first covariant derivative of $\mathbb{X}$, will contain the connection coefficient. As the second covariant derivative, the field strength, $\mathbb{F}$, may contain the connection coefficient and curvature tensor. As the third covariant derivative, the field source, $\mathbb{S}$, will contain the connection coefficient, curvature tensor, and curvature derivative. b) If the field potential, $\mathbb{A}$, is chosen as the first-rank tensor, the field strength, $\mathbb{F}$, as the first covariant derivative of $\mathbb{A}$, will contain the connection coefficient. As the second covariant derivative, the field source, $\mathbb{S}$, may contain the connection coefficient and curvature tensor. c) When the angular momentum, $\mathbb{L}$, is chosen as the first-rank tensor, the torque, $\mathbb{W}$, as the first covariant derivative of $\mathbb{L}$, will contain the connection coefficient. As the second covariant derivative, the force, $\mathbb{N}$, may contain the connection coefficient and curvature tensor. It should be noted that three different first-rank tensors in the above may not be in the same tangent frame or curved space.

In a similar way, the acceleration, $\mathbb{D}$, can be chosen as the first-rank tensor, and then the jerk, $\mathbb{K}$, as the first covariant derivative of $\mathbb{D}$, will contain the connection coefficient. As the second covariant derivative, the jounce, $\mathbb{J}$, may contain the connection coefficient and curvature tensor. Of course, as one first-rank tensor, the curved space relevant to the acceleration, $\mathbb{D}$, may be different from that relevant to these three first-rank tensors in the above.

Obviously the covariant derivative of a physics quantity may include a few parameters of the curved space. Comparing the physics quantity in the curved space with that in the flat space, we are able to distinguish different bending degrees of the curved space. Contrasting several sets of measured values of physics quantities will conclude the bending degree of a curved space.

\section{Complex-quaternion curved space}

Under an orthogonal transformation of the coordinate system with the complex quaternions, the norm of the complex quaternions remains unchanged. The norm (or the arc length) is identical to the space-time interval in the physics, and is able to be applied to describe the gravitational field equations in the complex quaternion curved space.

Apparently, in the complex quaternion curved space, the underlying space is the quaternion space, and the tangent space is the quaternion space also. Making use of the complex quaternion orthogonality and affine frame, it is able to define the quaternion metric and covariant derivative, studying the gravitational properties in the complex quaternion curved space.

\subsection{Quaternion metric}

In the complex quaternion flat space, the complex quaternion radius vector is, $\mathbb{H}_g = i \textbf{\emph{i}}_0 h^0 + \textbf{\emph{i}}_r h^r $, with the basis vector being ${\emph{\textbf{i}}_j}$. According to the multiplication of quaternion, the norm $S$ is written as, $S^2 = \mathbb{H}_g \circ \mathbb{H}_g^* = - (h^0)^2 + (h^1)^2 + (h^2)^2 + (h^3)^2$. Obviously the differential, $d S$, of the norm is able to be chosen as the arc length of the complex quaternion (rather than the real-quaternion) flat space. And that this norm accords with the requirement of space-time interval in the physics. Herein $i$ is the imaginary unit. $h^j$ is real. $\circ$ is the quaternion multiplication. $*$ is conjugate of quaternion. $\emph{\textbf{i}}_0^2 = 1$, and $\emph{\textbf{i}}_r^2 = - 1$. $\emph{\textbf{i}}_j^2 = \emph{\textbf{i}}_j \circ \emph{\textbf{i}}_j$, for each subscript $j$. For the superscript and subscript, there are, $j, k, m, n = 0, 1, 2, 3$; $r, q = 1, 2, 3$.

In the tangent space (that is, the locally flat space) of the complex quaternion curved space, the radius vector is, $\mathbb{H}_g = i \textbf{\emph{e}}_0 c^0 + \textbf{\emph{e}}_r c^r $ , with the tangent-frame quaternion being ${\textbf{e}_j}$ . In the orthogonal and unequal-length affine frame, the metric of complex quaternion curved space is defined as,
\begin{equation}
d S^2 = d \mathbb{H}_g^* \circ d \mathbb{H}_g  = g_{\overline{j}k} d \overline{u^j} d u^k   ~,
\end{equation}
where the metric coefficient, $g_{\overline{j}k} = \textbf{e}_j^* \circ \textbf{e}_k$ , is quaternion-Hermitian. The orthogonal tangent-frame quaternion is, $\textbf{e}_j = \partial \mathbb{H}_g / \partial u^j$ . $\textbf{e}_0$ is the scalar part, while $\textbf{e}_r$ is the component of vector part. $\textbf{e}_k$ is unequal-length, and does not contain the imaginary unit $i$ . $ u^0 = i c^0 $ , and $ u^r = c^r $ . $ ( u^j )^* = \overline{u^j} $ , and it denotes that the correlated tangent-frame component, $\textbf{e}_j$ , is quaternion-conjugate. $c^0 = v_0 t$, $v_0$ is the speed of light, and $t$ is the time. $c^j$ is real, and $g_{\overline{j}k}$ is scalar.

\subsection{Quaternion parallel translation}

The normal parallel translation (Euclidean space) and the Levi-Civita parallel translation (Riemannian space) both are not suitable for the complex quaternion curved space. One reason is that the spaces associated with in the physics are not only the complex quaternion flat space (tangent space) but also the pseudo-Riemannian space (curved space). Therefore it is necessary to introduce one new concept of parallel translation. It is similar to the Levi-Civita parallel translation, and is able to satisfy the requirement of the complex quaternion space and of the space-time interval.

In a complex quaternion space, when the quaternion product of two complex quaternions, $\mathbb{G} ( g^0 , g^1 , g^2 , g^3 )$ and $\mathbb{Z} ( z^0 , z^1 , z^2 , z^3 )$, equals to zero, that is, $\mathbb{G}^* \circ \mathbb{Z} = 0$, the two quaternions, $\mathbb{G}$ and $\mathbb{Z}$ , are perpendicular to each other. This definition is called as the quaternion orthogonality. After that we can define a quaternion connection and parallel translation in the complex quaternion curved space, referring to the inference procedure of Levi-Civita parallel translation in the Riemannian space. Herein $g^j$ and $z^j$ are all scalar.

In the complex quaternion curved space, the complex quaternion physics quantity $\mathbb{A}_1$ , in the tangent space $\mathbb{T}_1$ at a point $M_1$ on the complex quaternion manifold, can be decomposed in the tangent space $\mathbb{T}_2$ at the point $M_2$ near $M_1$ . According to the definition of quaternion orthogonality, the physical quantity $\mathbb{A}_1$ can be separated into the projection component $\mathbb{A}_2$ in $\mathbb{T}_2$ , and the orthogonal component $\mathbb{G}_2$ perpendicular to $\mathbb{T}_2$ . In case the differential, $\mathbb{A}_2 - \mathbb{A}_1 = 0$, the physical quantity $\mathbb{A}_2$ is the parallel translation of $\mathbb{A}_1$ . And this definition is called as the quaternion parallel translation. Especially, when the scalar parts of $\mathbb{A}_1$ and of $\mathbb{A}_2$ are all null, $\mathbb{A}_1$ and $\mathbb{A}_2$ both will be degenerated into the vectors. Further the orthogonality of quaternion will be reduced to that of vector, and the quaternion parallel translation to the Levi-Civita parallel translation.

\subsection{Quaternion covariant derivative}

In the Riemannian space, substituting the Levi-Civita parallel translation and tangent frame vector for the normal parallel translation and basis vector in the flat space respectively, the approach to define the partial derivative via the limit can be directly extended into the definition of covariant derivative of the Riemannian space.

Similarly in the complex quaternion curved space, substituting the quaternion parallel translation, tangent-frame quaternion, and quaternion orthogonality for the Levi-Civita parallel translation, tangent-frame vector, and vector orthogonality respectively, the covariant derivative of the pseudo-Riemannian space can be extended to that of the complex quaternion curved space directly.

For the first-rank contravariant tensor $Y^j$ of a point $M_2$ in the complex quaternion curved space, the component of quaternion covariant derivative with respect to the coordinate $u^k$ is,
\begin{equation}
\triangledown_k Y^n = \partial ( \delta_j^n Y^j ) / \partial u^k + \Gamma^n_{jk} Y^j   ~,
\end{equation}
where $\Gamma^n_{jk}$ is the connection coefficient. Making use of Eq.(1), one can deduce the connection coefficient from the above (Appendix B). $Y^n$ and $\Gamma^n_{jk}$ are all scalar.

\section{Gravitational field equations}

In the complex quaternion curved space, the gravitational potential had been transformed from the rectangular coordinate system (in the flat space) to the orthogonal and unequal-length affine frame (in the tangent space). The gravitational potential, $\mathbb{A}_g ( i a^0 , a^1 , a^2 , a^3 )$ , can be defined from the integrating function of gravitational potential, $\mathbb{X}_g$ ,
\begin{equation}
\mathbb{A}_g = i \lozenge^\star \circ \mathbb{X}_g  ~,
\end{equation}
where $\mathbb{A}_g = i \lozenge^\star \circledcirc \mathbb{X}_g + i \lozenge^\star \circledast \mathbb{X}_g$ . $i \lozenge^\star \circledcirc \mathbb{X}_g$ and $i \lozenge^\star \circledast \mathbb{X}_g$ denote the scalar and vector parts of $\mathbb{A}_g$ respectively. $\lozenge a^j = i \textbf{e}_0 \triangledown_0 a^j + \delta^{rq} \textbf{e}_r \triangledown_q a^j$. $\nabla a^j = \delta^{rq} \textbf{e}_r \triangledown_q a^j$. $\textbf{e}^j = g^{jk} \textbf{e}_k$. $\mathbb{X}_g = x^j \textbf{e}_j$. The gauge equation is chosen as, $\nabla \times (x^r \textbf{e}_r) = 0$. $\mathbb{A}_g = i a + \textbf{a}$. $a = a^0 \textbf{e}_0$ , and $\textbf{a} = a^r \textbf{e}_r$ . Apparently, in the affine frame, the gravitational potential includes not only the physics quantity but also the spatial parameter of the complex quaternion curved space (Table 1). The quaternion operator $\lozenge$ is extended from the flat space (see Ref.[9]) to the curved space in this paper. And the quaternion operator is different from the Dirac operator. The former comes from the algebra of quaternions, while the latter derived from the mass-energy relation in the Special Theory of Relativity, which is applied to the Quantum Mechanics. $\star$ is the complex conjugate. $a^j$ is real.

The gravitational strength, $\mathbb{F}_g ( i f^0 , f^1 , f^2 , f^3 )$ , is defined as
\begin{equation}
\mathbb{F}_g = \lozenge \circ \mathbb{A}_g  ~  ,
\end{equation}
where $\mathbb{F}_g = \lozenge \circledcirc \mathbb{A}_g + \lozenge \circledast \mathbb{A}_g $ . The scalar part of $\mathbb{F}_g$ is, $\lozenge \circledcirc \mathbb{A}_g = i f^0 \textbf{e}_0$, and the vector part of $\mathbb{F}_g$ is, $\lozenge \circledast \mathbb{A}_g = f^r \textbf{e}_r$. The gauge equation of gravitational potential is chosen as, $f^0 = 0$. The vector part of gravitational strength can be separated into two components, $f^r \textbf{e}_r = i \textbf{g} / v_0 + \textbf{b}$ . One component, $\textbf{g} / v_0 = (\textbf{e}_0 \triangledown_0) \circ \textbf{a} + \nabla \circ a$, is relevant to the acceleration, while the other, $\textbf{b} = \nabla \times \textbf{a}$ , is associated to the precession angular velocity. $f^0$ is real, and $f^r$ is a complex number.

The gravitational source, $\mathbb{S}_g ( i s^0 , s^1 , s^2 , s^3 )$ , is defined as,
\begin{equation}
- \mu \mathbb{S} = - ( \mu_g \mathbb{S}_g - i \mathbb{F}_g^* \circ \mathbb{F}_g / v_0) = ( \lozenge + i \mathbb{F}_g / v_0)^* \circ \mathbb{F}_g   ~,
\end{equation}
or
\begin{equation}
- \mu_g \mathbb{S}_g = \lozenge^* \circ \mathbb{F}_g  ~,
\end{equation}
where $- \mu_g \mathbb{S}_g = \lozenge^* \circledcirc \mathbb{F}_g +  \lozenge^* \circledast \mathbb{F}_g $ . The scalar part of $\mathbb{S}_g$ is, $- \lozenge^* \circledcirc \mathbb{F}_g / \mu_g = i s^0 \textbf{e}_0$ , which is associated to the mass density. And the vector part of $\mathbb{S}_g$ is, $- \lozenge^* \circledast \mathbb{F}_g / \mu_g = s^r \textbf{e}_r$ , which is relevant to the density of linear momentum. $\mu$ and $\mu_g$ both are the coefficients. $s^j$ is real.

Considering the complex coordinate and definition of gravitational strength, it is possible to expand the above to achieve the four component equations of gravitational field in the complex quaternion curved space (Appendix C). Now the gravitational field equations deal with some spatial parameters of the complex quaternion curved space. For the first-rank contravariant tensor $Y^n$ in the complex quaternion curved space, the complex quaternion physics quantity $\lozenge Y^n$ is associated to the connection coefficient $\Gamma^n_{jk}$ . Meanwhile the complex quaternion physics quantities, $\lozenge^* \circ (\lozenge Y^n)$ and $\lozenge \circ (\lozenge^* Y^n)$, contain the connection coefficient $\Gamma^n_{jk}$ and the curvature $R_{j\overline{k}m}^{~~~~n}$ . As a result, choosing the gravitational potential as the first-rank contravariant tensor, the gravitational strength will be relevant to the connection coefficient $\Gamma^n_{jk}$ , while the gravitational source will be associated to the connection coefficient $\Gamma^n_{jk}$ and the curvature $R_{j\overline{k}m}^{~~~~n}$ .

In the tangent space, in order to facilitate comparison among the field equations in different coordinate systems, the gravitational field equations in the above may be firstly transformed from the orthogonal and unequal-length coordinate system into the orthogonal and equal-length coordinate system. Subsequently comparing the gravitational field equations in the orthogonal and equal-length coordinate system with that in the flat space, it is possible to appraise the bending degree of the complex quaternion curved space.

Certainly the deduction approach of the gravitational field equations in the complex quaternion curved space can be used as a reference to be extended to that of electromagnetic field equations.

\begin{table}[h]
\caption{The multiplication of the operator with the physics quantity of gravitational and electromagnetic fields, in the complex octonion curved space.}
\label{tab:table1}
\centering
\begin{tabular}{ll}
\hline\hline
definition                                                      &   expression~meaning                                                                                \\
\hline
$\nabla \cdot (\textbf{e}_r a^r)$                               &  $(\textbf{e}_1 \cdot \textbf{e}_1) \triangledown_1 a^1
                                                                         + (\textbf{e}_2 \cdot \textbf{e}_2) \triangledown_2 a^2
                                                                         + (\textbf{e}_3 \cdot \textbf{e}_3) \triangledown_3 a^3 $                                    \\
$\nabla \times (\textbf{e}_r a^r)$                              &  $(\textbf{e}_2 \times \textbf{e}_3) ( \triangledown_2 a^3 - \triangledown_3 a^2 )
                                                                         + (\textbf{e}_3 \times \textbf{e}_1) ( \triangledown_3 a^1 - \triangledown_1 a^3 )$          \\
$$                                                              &  ~~~ $ + (\textbf{e}_1 \times \textbf{e}_2) ( \triangledown_1 a^2 - \triangledown_2 a^1 )$          \\
$\nabla \circ (\textbf{e}_0 a^0) $                              &  $(\textbf{e}_1 \circ \textbf{e}_0) \triangledown_1 a^0
                                                                         + (\textbf{e}_2 \circ \textbf{e}_0) \triangledown_2 a^0
                                                                         + (\textbf{e}_3 \circ \textbf{e}_0) \triangledown_3 a^0  $                                   \\
$(\textbf{e}_0 \triangledown_0) \circ (\textbf{e}_r a^r)$       &  $(\textbf{e}_0 \circ \textbf{e}_1) \triangledown_0 a^1
                                                                         + (\textbf{e}_0 \circ \textbf{e}_2) \triangledown_0 a^2
                                                                         + (\textbf{e}_0 \circ \textbf{e}_3) \triangledown_0 a^3  $                                   \\
$\nabla \cdot (\textbf{E}_r A^r)$                               &  $ (\textbf{e}_1 \cdot \textbf{E}_1) \triangledown_1 A^1
                                                                         + (\textbf{e}_2 \cdot \textbf{E}_2) \triangledown_2 A^2
                                                                         + (\textbf{e}_3 \cdot \textbf{E}_3) \triangledown_3 A^3 $                                    \\
$\nabla \times (\textbf{E}_r A^r)$                              &  $ (\textbf{e}_2 \times \textbf{E}_3) ( \triangledown_2 A^3 - \triangledown_3 A^2 )
                                                                         + (\textbf{e}_3 \times \textbf{E}_1) ( \triangledown_3 A^1 - \triangledown_1 A^3 )$          \\
$$                                                              &  ~~~ $ + (\textbf{e}_1 \times \textbf{E}_2) ( \triangledown_1 A^2 - \triangledown_2 A^1 )$          \\
$\nabla \circ (\textbf{E}_0 A^0)$                               &  $ (\textbf{e}_1 \circ \textbf{E}_0) \triangledown_1 A^0
                                                                         + (\textbf{e}_2 \circ \textbf{E}_0) \triangledown_2 A^0
                                                                         + (\textbf{e}_3 \circ \textbf{E}_0) \triangledown_3 A^0  $                                   \\
$(\textbf{e}_0 \triangledown_0) \circ (\textbf{E}_r A^r)$       &  $ (\textbf{e}_0 \circ \textbf{E}_1) \triangledown_0 A^1
                                                                         + (\textbf{e}_0 \circ \textbf{E}_2) \triangledown_0 A^2
                                                                         + (\textbf{e}_0 \circ \textbf{E}_3) \triangledown_0 A^3  $                                   \\
\hline\hline
\end{tabular}
\end{table}

\section{Complex-octonion curved space}

Under an orthogonal transformation of the coordinate system with the complex octonions, the norms of the complex octonions remain unchanged. And the norm includes the space-time interval (or the arc length) in the physics, and is possible to be applied to describe simultaneously the gravitational field equations and the electromagnetic field equations in the complex octonion curved space. By means of the complex octonion orthogonality and affine frame, it is possible to define the octonion metric and covariant derivative, depicting the gravitational and electromagnetic properties in the complex octonion curved space.

What needs to be explained specially is, that the gravitational field involved in is the quaternion operator and physics quantity in the complex quaternion space. Meanwhile what the electromagnetic field concerned with is, the physics quantity in the complex $S$-quaternion space, and the quaternion operator in the complex quaternion space. Consequently the paper deals with one comparatively simple situation. That is, the paper will be only involved in the interval (or the arc length) of the complex quaternion space.

In the complex octonion curved space, the underlying space is the octonion space, and the tangent space is the octonion space also. A tangent frame system (complex quaternion) can discuss the gravitational field, while the other tangent frame system (complex $S$-quaternion) may research the electromagnetic field, in the curved space.

Some calculations in the context may be necessary to be dealt with the non-associativity of the octonions, for instance, some multiplications in the Appendices B, D, and Table 3. One can multiply the next argument by an existing octonion from the left-side, step by step, to ensure the existence and uniqueness of the product relevant to physical quantities.

\subsection{Octonion metric}

In the complex quaternion flat space, the complex quaternion radius vector is, $\mathbb{H}_g = i \textbf{\emph{i}}_0 h^0 + \textbf{\emph{i}}_r h^r $, with the basis vector being ${\emph{\textbf{i}}_j}$ . Meanwhile, in the complex $S$-quaternion flat space for the electromagnetic field, the complex $S$-quaternion radius vector is, $\mathbb{H}_e =  i \textbf{\emph{I}}_0 H^0 + \textbf{\emph{I}}_r H^r $ , with the basis vector being ${\emph{\textbf{I}}_j}$ . In the complex octonion flat space, two radius vectors, $\mathbb{H}_g$ and $\mathbb{H}_e$ , can be combined together to become one complex octonion radius vector,
\begin{equation}
\mathbb{H} (h^\alpha) = \mathbb{H}_g + k_{eg} \mathbb{H}_e  = i h^0 \emph{\textbf{i}}_0 + h^r \emph{\textbf{i}}_r + i h^4 \emph{\textbf{i}}_4 + h^{4+r} \emph{\textbf{i}}_{4+r}   ~,
\end{equation}
where $h^{j+4} = k_{eg} H^j$ , $\emph{\textbf{i}}_{j+4} = \emph{\textbf{I}}_j$ . $\mathbb{H}_g = i h^0 \emph{\textbf{i}}_0 + h^r \emph{\textbf{i}}_r$ , $\mathbb{H}_e = i H^0 \emph{\textbf{I}}_0 + H^r \emph{\textbf{I}}_r$ . The coefficient $k_{eg}$ meets the demand for the dimensional homogeneity in the physics, according to the Appendix A. $h^j$ and $H^j$ are all real. $\emph{\textbf{i}}_0^2 = 1$, $\emph{\textbf{i}}_v^2 = - 1$. $\emph{\textbf{i}}_\alpha^2 = \emph{\textbf{i}}_\alpha \circ \emph{\textbf{i}}_\alpha$, for each subscript $\alpha$. $\alpha, \beta, \gamma, \lambda = 0, 1, 2, 3, 4, 5, 6, 7$. $v = 1, 2, 3, 4, 5, 6, 7$.

According to the multiplication of octonion, the norm $S$ is written as, $S^2 = \mathbb{H} \circ \mathbb{H}^*$. The differential, $d S$, of the norm is able to be chosen as the arc length of the complex octonion (rather than the real-octonion) curved space. Obviously, in case the contribution of $k_{eg} \mathbb{H}_e$ can be neglected, this norm will accord with the requirement of space-time interval in the physics. Herein $\circ$ and $*$ are upgraded to the octonion multiplication and conjugate respectively.

In the complex quaternion curved space for the gravitational field, the complex quaternion radius vector is, $\mathbb{H}_g = i \textbf{\emph{e}}_0 c^0 + \textbf{\emph{e}}_r c^r $, with the tangent-frame quaternion being ${\textbf{e}_j}$. In the complex $S$-quaternion curved space for the electromagnetic field, the complex $S$-quaternion radius vector is, $\mathbb{H}_e =  i \textbf{\emph{E}}_0 C^0 + \textbf{\emph{E}}_r C^r $ , with the tangent-frame $S$-quaternion being ${\textbf{E}_j}$. Therefore, in the tangent space of complex octonion curved space, the complex octonion radius vector is, $\mathbb{H} = \mathbb{H}_g + k_{eg} \mathbb{H}_e$. Making use of the substitution, $c^{j+4} = k_{eg} C^j$ and $\textbf{e}_{j+4} = \textbf{E}_j$ , the complex octonion radius vector can be rewritten as, $\mathbb{H} (c^\alpha) = i c^0 \textbf{e}_0 + c^r \textbf{e}_r + i c^4 \textbf{e}_4 + c^{4+r} \textbf{e}_{4+r}$ . Herein $c^j$ and $C^j$ are all real.

In the orthogonal and unequal-length affine frame, the metric of complex octonion curved space is defined as,
\begin{equation}
d S^2 = d \mathbb{H}^* \circ d \mathbb{H}  = g_{\overline{\alpha} \beta} d \overline{u^\alpha} d u^\beta   ~,
\end{equation}
where the metric coefficient, $g_{\overline{\alpha} \beta} = \textbf{e}_\alpha^* \circ \textbf{e}_\beta$ , is octonion-Hermitian. The orthogonal tangent-frame octonion is, $\textbf{e}_\alpha = \partial \mathbb{H} / \partial u^\alpha$ . $\textbf{e}_0$ is the scalar part, while $\textbf{e}_v$ is the component of vector part. $\textbf{e}_\beta$ is unequal-length, and does not contain the imaginary unit $i$ . $u^0 = i c^0$ , $u^r = c^r$ , $u^4 = i c^4$ , and $u^{4+r} = c^{4+r}$ . $ ( u^\alpha )^* = \overline{u^\alpha} $ , and it denotes that the correlated tangent-frame component, $\textbf{e}_\alpha$ , is octonion-conjugate. $g_{\overline{\alpha} \beta}$ is scalar.

However, in the complex octonion curved space, what the paper will be involved in is, the mathematical manipulation between the quaternion operator (rather than the octonion operator) in the complex quaternion space, with the physics quantity of electromagnetic field in the complex $S$-quaternion space. So that the discussion of the space-time interval in the paper will be constrained to deal only with the component of the complex quaternion radius vector, $\mathbb{H}_g$ .

\begin{table}[h]
\caption{Comparison of major characteristics in some flat and curved spaces, including the pseudo-Riemannian space, quaternion space, and octonion space.}
\label{tab:table2}
\centering
\begin{tabular}{lllll}
\hline \hline
Terms                    &    pseudo-                 &    Quaternion                &     Octonion                         \\
                         &    Riemannian space        &    space                     &     space                            \\
\hline
tangent space            &    vector space            &    quaternion space          &     octonion space                   \\
orthogonality            &    vector                  &    quaternion                &     octonion                         \\
parallel translation     &    Levi-Civita             &    quaternion                &     octonion                         \\
tangent frame            &    vector/scalar           &    quaternion                &     octonion                         \\
metric                   &    scalar product          &    scalar product            &     scalar product                   \\
                         &                            &    ~~ of quaternions         &     ~~ of octonions                  \\
connection coefficient   &    $\Gamma^n_{jk}$         &    $\Gamma^n_{jk}$           &     $\Gamma^\beta_{\alpha \gamma}$   \\
covariant derivative     &    $\nabla_k A^j$          &    $\nabla_k A^j$            &     $\nabla_\gamma A^\beta$          \\
\hline \hline
\end{tabular}
\end{table}

\subsection{Octonion parallel translation}

In the complex octonion curved space, they are not suitable enough that the normal parallel translation (Euclidean space), the Levi-Civita parallel translation (Riemannian space), and the quaternion parallel translation (quaternion curved space). One of reasons is that the spaces in the physics associate with not only the complex octonion flat space (tangent space) but also the pseudo-Riemannian space (curved space). Subsequently it is necessary to introduce one new concept of parallel translation. It is similar to the quaternion parallel translation, and is able to meet the demand for the complex octonion space and of the space-time interval.

In one complex octonion space, when the product of two complex octonions, $\mathbb{G} ( g^\alpha )$ and $\mathbb{Z} ( z^\beta )$ , is equal to zero, that is, $\mathbb{G}^* \circ \mathbb{Z} = 0$ , two octonions, $\mathbb{G} (g^\alpha)$ and $\mathbb{Z} (z^\beta)$ , are perpendicular to each other. This definition is called as the octonion orthogonality. Therefore it is able to define an octonion connection and parallel translation in the complex octonion curved space, referring to the inference procedure of quaternion parallel translation in the complex quaternion curved space. Herein $g^\alpha$ and $z^\beta$ are all scalar.

In the complex octonion curved space, the complex octonion physics quantity $\mathbb{A}_1$ , in the tangent space $\mathbb{T}_1$ of one point $M_1$ on the complex octonion manifold, can be decomposed in the tangent space $\mathbb{T}_2$ of the point $M_2$ near $M_1$. According to the definition of octonion orthogonality, $\mathbb{A}_1$ can be separated into the projection component $\mathbb{A}_2$ in $\mathbb{T}_2$ , and the orthogonal component $\mathbb{G}_2$ perpendicular to $\mathbb{T}_2$ . In case the differential, $\mathbb{A}_2 - \mathbb{A}_1 = 0$, the physical quantity $\mathbb{A}_2$ is the parallel translation of $\mathbb{A}_1$ . And this definition is called as the octonion parallel translation. Especially, when the scalar parts of $\mathbb{A}_1$ and of $\mathbb{A}_2$ are all null, $\mathbb{A}_1$ and $\mathbb{A}_2$ both will be degenerated into the vectors. Further the orthogonality of octonion is reduced to that of vector, and the octonion parallel translation to the Levi-Civita parallel translation. When the complex octonions are reduced to the complex quaternions, the orthogonality of octonion is degenerated into that of quaternion, and the octonion parallel translation into the quaternion parallel translation (Table 2).

\subsection{Octonion covariant derivative}

In the complex octonion curved space, substituting the octonion parallel translation, tangent-frame octonion, and octonion orthogonality for the quaternion parallel translation, tangent-frame quaternion, and quaternion orthogonality respectively, the covariant derivative of the complex quaternion curved space can be extended into that of the complex octonion curved space directly.

For the first-rank contravariant tensor $Y^\beta$ of a point $M_2$ in the complex octonion curved space, the component of octonion covariant derivative with respect to the coordinate $u^\gamma$ is,
\begin{equation}
\triangledown_\gamma Y^\beta = \partial ( \delta_\alpha^\beta Y^\alpha ) / \partial u^\gamma + \Gamma^\beta_{\alpha \gamma} Y^\alpha   ~ ,
\end{equation}
where $\Gamma^\beta_{\alpha \gamma}$ is the connection coefficient. One can deduce the connection coefficient, from the Appendix B. $Y^\beta$ and $\Gamma^\beta_{\alpha \gamma}$ are all scalar.

\section{Electromagnetic field equations}

In the complex octonion curved space, the electromagnetic potential and gravitational potential had been transformed from the rectangular coordinate system (flat space) to the orthogonal and unequal-length affine frame (tangent space). From the octonion integrating function, $\mathbb{X}$, of field potential, the octonion field potential, $\mathbb{A} = \mathbb{A}_g + k_{eg} \mathbb{A}_e$ , is defined as,
\begin{equation}
\mathbb{A} = i \lozenge^\star \circ \mathbb{X}   ~,
\end{equation}
where $\mathbb{A} = i \lozenge^\star \circledcirc \mathbb{X} + i \lozenge^\star \circledast \mathbb{X}$ . $\mathbb{X} = \mathbb{X}_g + k_{eg} \mathbb{X}_e$ . The integrating function, $\mathbb{X}_e$, of electromagnetic potential is one $S$-quaternion physics quantity, and $\mathbb{X}_e = X^j \textbf{E}_j$. The electromagnetic potential is $\mathbb{A}_e ( i A^0 , A^1 , A^2 , A^3 ) = i \lozenge^\star \circ \mathbb{X}_e$ , with $\mathbb{A}_e = i \lozenge^\star \circledcirc \mathbb{X}_e + i \lozenge^\star \circledast \mathbb{X}_e$ . $i \lozenge^\star \circledcirc \mathbb{X}_e$ and $i \lozenge^\star \circledast \mathbb{X}_e$ denote respectively the `scalar' and vector parts of $\mathbb{A}_e$ . The gauge equation is chosen as, $\nabla \times (X^r \textbf{E}_r) = 0$. $\mathbb{A}_e = i \textbf{A}_Q + \textbf{A}$. $\textbf{A}_Q = A^0 \textbf{E}_0$, and $\textbf{A} = A^r \textbf{E}_r$. Apparently, in the affine frame, the electromagnetic potential includes not only the physics quantity (in the complex $S$-quaternion curved space) but also the spatial parameter of curved space (in the complex quaternion curved space). $a^j$ and $A^j$ are all real.

The octonion field strength, $\mathbb{F} = \mathbb{F}_g + k_{eg} \mathbb{F}_e$ , is defined as,
\begin{equation}
\mathbb{F} = \lozenge \circ \mathbb{A}   ~ ,
\end{equation}
where $\mathbb{F} = \lozenge \circledcirc \mathbb{A} + \lozenge \circledast \mathbb{A} $ . The electromagnetic strength is, $\mathbb{F}_e ( i F^0 , F^1 , F^2 , F^3 ) = \lozenge \circ \mathbb{A}_e$. $\mathbb{F}_e = \lozenge \circledcirc \mathbb{A}_e + \lozenge \circledast \mathbb{A}_e $. The `scalar' part of $\mathbb{F}_e$ is, $\lozenge \circledcirc \mathbb{A} = i F^0 \textbf{E}_0$ , and the vector part of $\mathbb{F}_e$ is, $\lozenge \circledast \mathbb{A}_e = F^r \textbf{E}_r$ . The gauge equation of electromagnetic potential is chosen as, $F^0 = 0$. The vector part of electromagnetic strength can be separated into two components, $F^r \textbf{E}_r = i \textbf{E} / v_0 + \textbf{B}$. One component, $\textbf{E} / v_0 = (\textbf{e}_0 \triangledown_0) \circ \textbf{A} + \nabla \circ \textbf{A}_Q$, is the electric field intensity, while the other, $\textbf{B} = \nabla \times \textbf{A}$, is the magnetic flux density. $F^0$ is real, and $F^r$ is one complex number.

The octonion field source, $\mu \mathbb{S} = \mu_g \mathbb{S}_g + k_{eg} \mu_e \mathbb{S}_e$ , is defined as,
\begin{equation}
- \mu \mathbb{S}  = - ( \mu_g \mathbb{S}_g + k_{eg} \mu_e \mathbb{S}_e - i \mathbb{F}^* \circ \mathbb{F} / v_0)  = ( \lozenge + i \mathbb{F} / v_0)^* \circ \mathbb{F}   ~,
\end{equation}
or
\begin{equation}
- \mu_g \mathbb{S}_g = \lozenge^* \circ \mathbb{F}_g    ~,  ~~~ - \mu_e \mathbb{S}_e = \lozenge^* \circ \mathbb{F}_e   ~,
\end{equation}
where $- \mu_g \mathbb{S}_g = \lozenge^* \circledcirc \mathbb{F}_g + \lozenge^* \circledast \mathbb{F}_g$ , while $- \mu_e \mathbb{S}_e = \lozenge^* \circledcirc \mathbb{F}_e + \lozenge^* \circledast \mathbb{F}_e$ . The electromagnetic source is $\mathbb{S}_e ( i S^0 , S^1 , S^2 , S^3 )$. The `scalar' part of $\mathbb{S}_e$ is, $- \lozenge^* \circledcirc \mathbb{F}_e / \mu_e = i S^0 \textbf{E}_0$, and is associated to the density of electric charge. And the vector part of $\mathbb{S}_e$ is, $- \lozenge^* \circledast \mathbb{F}_e / \mu_e = S^r \textbf{E}_r$ , and is relevant to the density of electric current. $\mu$ and $\mu_e$ are the coefficients. $S^j$ is real.

Considering the complex coordinate and definition of the electromagnetic strength, it is able to expand the above to achieve the four component equations of electromagnetic field in the complex $S$-quaternion curved space (Appendix D). Therefore the field equations deal with some spatial parameters of the complex octonion curved space. For the first-rank contravariant tensor $Y^\beta$ in the complex octonion curved space, the complex octonion physics quantity $\lozenge Y^\beta$ is associated to the connection coefficient $\Gamma^\beta_{\alpha \gamma}$ , while the complex octonion physics quantities,  $\lozenge^* \circ (\lozenge  Y^\beta)$ and $\lozenge \circ (\lozenge^* Y^\beta)$ , are relevant to the connection coefficient $\Gamma^\beta_{\alpha \gamma}$ and the curvature $R_{\alpha \overline{\gamma} \lambda}^{~~~~\beta}$ . As a result, choosing the electromagnetic (or gravitational) potential as the first-rank contravariant tensor, the electromagnetic (or gravitational) strength will be relevant to the connection coefficient $\Gamma^\beta_{\alpha \gamma}$, while the electromagnetic (or gravitational) source will be associated to the connection coefficient $\Gamma^\beta_{\alpha \gamma}$ and curvature $R_{\alpha \overline{\gamma} \lambda}^{~~~~\beta}$ .

In the $S$-quaternion tangent space, in order to facilitate comparison among the electromagnetic field equations in different coordinate systems, the electromagnetic field equations in the above may be firstly transformed from the orthogonal and unequal-length coordinate system into the orthogonal and equal-length coordinate system. Subsequently comparing the electromagnetic field equations in the orthogonal and equal-length coordinate system with that in the flat space, it is able to estimate the bending degree of the complex $S$-quaternion curved space. In the process of providing a contrast for the electromagnetic field equations, it is necessary to take into account the influence of the bending degree of complex quaternion curved space on the electromagnetic field equations. As a result, it is capable of measuring the bending degree of the complex octonion curved space, making use of contrasting the octonion field equations in the curved space with that in the flat space.

\begin{table}[h]
\caption{Some definitions of the physics quantity relevant to the gravitational and electromagnetic fields in the complex octonion curved space.}
\label{tab:table3}
\centering
\begin{tabular}{lll}
\hline\hline
physics~quantity             &   definition                                                                                 \\
\hline
radius~vector                &  $\mathbb{H} = \mathbb{H}_g + k_{eg} \mathbb{H}_e  $                                         \\
integral~function            &  $\mathbb{X} = \mathbb{X}_g + k_{eg} \mathbb{X}_e  $                                         \\
field~potential              &  $\mathbb{A} = i \lozenge^\star \circ \mathbb{X}  $                                          \\
field~strength               &  $\mathbb{F} = \lozenge \circ \mathbb{A}  $                                                  \\
field~source                 &  $\mu \mathbb{S} = - ( i \mathbb{F} / v_0 + \lozenge )^* \circ \mathbb{F} $                  \\
linear~momentum              &  $\mathbb{P} = \mu \mathbb{S} / \mu_g $                                                      \\
angular~momentum             &  $\mathbb{L} = ( \mathbb{H} + k_{rx} \mathbb{X} )^\star \circ \mathbb{P} $                   \\
octonion~torque              &  $\mathbb{W} = - v_0 ( i \mathbb{F} / v_0 + \lozenge ) \circ \mathbb{L} $                    \\
octonion~force               &  $\mathbb{N} = - ( i \mathbb{F} / v_0 + \lozenge ) \circ \mathbb{W} $                        \\
\hline\hline
\end{tabular}
\end{table}

\section{Octonion angular momentum}

In the complex octonion curved space for the gravitational and electromagnetic fields, the octonion linear momentum, $\mathbb{P} ( p^j , P^j ) = \mathbb{P}_g + k_{eg} \mathbb{P}_e$, is defined from the octonion field source,
\begin{equation}
\mathbb{P} = \mu \mathbb{S} / \mu_g ~,
\end{equation}
where the component of the octonion linear momentum, $\mathbb{P}$ , in the complex quaternion space is, $\mathbb{P}_g = \{ \mu_g \mathbb{S}_g - ( i \mathbb{F} / v_0 )^* \circ \mathbb{F} \} / \mu_g = i p^0 \textbf{e}_0 + p^q \textbf{e}_q$ . And the component of the octonion linear momentum, $\mathbb{P}$ , in the complex $S$-quaternion space is, $\mathbb{P}_e = \mu_e \mathbb{S}_e / \mu_g = i P^0 \textbf{E}_0 + P^q \textbf{E}_q$. $p^j$ and $P^j$ are all real.

Subsequently, in the complex octonion curved space, the octonion angular momentum, $\mathbb{L} ( l^j , L^j ) = \mathbb{L}_g + k_{eg} \mathbb{L}_e$, can be defined from the octonion linear momentum and radius vector,
\begin{equation}
\mathbb{L} = ( \mathbb{H} + k_{rx} \mathbb{X} )^\star \circ \mathbb{P}  ~,
\end{equation}
where $\mathbb{L} = ( \mathbb{H} + k_{rx} \mathbb{X} )^\star \circledcirc \mathbb{P} + ( \mathbb{H} + k_{rx} \mathbb{X} )^\star \circledast \mathbb{P}$ . The component of the octonion angular momentum, $\mathbb{L}$ , in the complex quaternion space is, $\mathbb{L}_g = ( \mathbb{H}_g + k_{rx} \mathbb{X}_g )^\star \circ \mathbb{P}_g + k_{eg}^2 ( \mathbb{H}_e + k_{rx} \mathbb{X}_e )^\star \circ \mathbb{P}_e = l^0 \textbf{e}_0 + l^q \textbf{e}_q$ . And the component of the octonion angular momentum, $\mathbb{L}$ , in the complex $S$-quaternion space is, $\mathbb{L}_e = ( \mathbb{H}_g + k_{rx} \mathbb{X}_g )^\star \circ \mathbb{P}_e + ( \mathbb{H}_e + k_{rx} \mathbb{X}_e )^\star \circ \mathbb{P}_g = L^0 \textbf{E}_0 + L^q \textbf{E}_q $. $l^j$ and $L^j$ are complex numbers.

In the above, the term, $( \mathbb{H} + k_{rx} \mathbb{X} )^\star \circledcirc \mathbb{P} = l^0 \textbf{e}_0 $, denotes the scalar part of the octonion angular momentum, $\mathbb{L}$ . While the term, $( \mathbb{H} + k_{rx} \mathbb{X} )^\star \circledast \mathbb{P} =  l^q \textbf{e}_q + k_{eg} ( L^0 \textbf{E}_0 + L^q \textbf{E}_q )$, indicates the vector part of the octonion angular momentum, $\mathbb{L}$. The real part of the term, $l^q \textbf{e}_q$ , is relevant to the angular momentum. The term, $L^0 \textbf{E}_0$ , is associated with the `scalar' magnetic moment. For the term, $L^q \textbf{E}_q$ , its imaginary part is connected with the electric dipole moment, while its real part is connected with the magnetic dipole moment.

\section{Octonion torque}

In the complex octonion curved space, the octonion torque, $\mathbb{W} ( w^j , W^j ) = \mathbb{W}_g + k_{eg} \mathbb{W}_e$ , is defined from the octonion linear momentum,
\begin{equation}
\mathbb{W} = - v_0 ( \lozenge + i \mathbb{F} / v_0) \circ \mathbb{L}  ~,
\end{equation}
where $\mathbb{W} = v_0 ( \lozenge + i \mathbb{F} / v_0) \circledcirc \mathbb{L} + v_0 ( \lozenge + i \mathbb{F} / v_0) \circledast \mathbb{L}$ . The component of the octonion torque, $\mathbb{W}$ , in the complex quaternion space is, $\mathbb{W}_g = - ( i \mathbb{F}_g \circ \mathbb{L}_g + i k_{eg}^2 \mathbb{F}_e \circ \mathbb{L}_e + v_0 \lozenge \circ \mathbb{L}_g ) = w^0 \textbf{e}_0 + w^q \textbf{e}_q$ . And the component of the octonion torque, $\mathbb{W}$, in the complex $S$-quaternion space is, $\mathbb{W}_e = - ( i \mathbb{F}_g \circ \mathbb{L}_e + i \mathbb{F}_e \circ \mathbb{L}_g + v_0 \lozenge \circ \mathbb{L}_e ) = W^0 \textbf{E}_0 + W^q \textbf{E}_q$. $w^j$ and $W^j$ are complex numbers.

In the above, the term, $v_0 ( \lozenge + i \mathbb{F} / v_0) \circledcirc \mathbb{L} = w^0 \textbf{e}_0$, denotes the scalar part of the octonion torque, $\mathbb{W}$ . And the term, $v_0 ( \lozenge + i \mathbb{F} / v_0) \circledast \mathbb{L} = w^q \textbf{e}_q + k_{eg} (W^0 \textbf{E}_0 + W^q \textbf{E}_q)$, indicates the vector part of the octonion torque, $\mathbb{W}$ . For the term, $w^0 \textbf{e}_0$ , its imaginary part is connected with the energy, while its real part is connected with the divergence of the angular momentum. Meanwhile for the term, $w^q \textbf{e}_q$, its imaginary part is associated with the force, and its real part is associated with the curl of the angular momentum. The real part of the term, $W^0 \textbf{E}_0$ , is relevant to the divergence of the magnetic dipole moment. And the part of the term, $W^q \textbf{E}_q$ , is dealt with the curl of the magnetic dipole moment and the derivative of the electric dipole moment and so on.

In the complex octonion curved space, when the octonion field potential is chosen as the first-rank tensor, the octonion field source, linear momentum, and angular momentum will be involved in the connection coefficient, curvature, and other spatial parameters of the complex octonion curved space. It means that the curved space will act on the octonion torque, including the energy, the torque, the divergence and curl of angular momentum, the divergence and curl of magnetic dipole moment, and the derivative of the electric dipole moment.

\section{Octonion force}

In the complex octonion curved space, the octonion force, $\mathbb{N} ( n^j , N^j ) = \mathbb{N}_g + k_{eg} \mathbb{N}_e$, is defined from the octonion torque,
\begin{equation}
\mathbb{N} = - ( \lozenge + i \mathbb{F} / v_0) \circ \mathbb{W}   ~,
\end{equation}
where $\mathbb{N}  = ( \lozenge + i \mathbb{F} / v_0) \circledcirc \mathbb{W} + ( \lozenge + i \mathbb{F} / v_0) \circledast \mathbb{W}  $ .

For the octonion force, the component, $\mathbb{N}_g = n^0 \textbf{e}_0 + n^q \textbf{e}_q $ , in the complex quaternion space is,
\begin{equation}
\mathbb{N}_g =  - ( i \mathbb{F}_g \circ \mathbb{W}_g / v_0 + \lozenge \circ \mathbb{W}_g + i k_{eg}^2 \mathbb{F}_e \circ \mathbb{W}_e / v_0 ) ~,
\end{equation}
and the component, $\mathbb{N}_e = N^0 \textbf{E}_0 + N^q \textbf{E}_q $ , in the complex $S$-quaternion space is,
\begin{equation}
\mathbb{N}_e =  - ( i \mathbb{F}_g \circ \mathbb{W}_e / v_0 + \lozenge \circ \mathbb{W}_e + i \mathbb{F}_e \circ \mathbb{W}_g / v_0  ) ~,
\end{equation}
where $n^j$ and $N^j$ are complex numbers.

In the above, the term, $( \lozenge + i \mathbb{F} / v_0) \circledcirc \mathbb{W} = n^0 \textbf{e}_0$ , denotes the scalar part of the octonion force, $\mathbb{N}$. And the term, $( \lozenge + i \mathbb{F} / v_0) \circledast \mathbb{W} = n^q \textbf{e}_q + k_{eg} (N^0 \textbf{E}_0 + N^q \textbf{E}_q)$, indicates the vector part of the octonion force, $\mathbb{N}$ . The real part of the term, $n^0 \textbf{e}_0$ , is relevant to the mass continuity equation, while the real part of the term, $N^0 \textbf{E}_0$, is associated with the current continuity equation. For the term, $n^q \textbf{e}_q$ , its imaginary part is connected with the force, and corresponded with the linear acceleration. And that its real part is corresponded with the precession angular velocity.

In the complex octonion curved space, the imaginary part of $(n^r \textbf{e}_r) / 2$ is the force, in the gravitational and electromagnetic fields, that is,
\begin{equation}
\textbf{N} = Im \{ (n^r \textbf{e}_r) / 2 \}  ~ ,
\end{equation}
where the force $\textbf{N}$ includes the inertial force, gravitational force, electromagnetic force, energy gradient force, and additional force term caused by the curved space. The additional force term is relevant to the connection coefficient and curvature and so forth of the complex octonion curved space.

A majority of force terms can be written approximately as,
\begin{eqnarray}
\textbf{N}_M \approx && - \nabla \circ (p^0 v_0 \textbf{e}_0) - ( \textbf{e}_0 \triangledown_0 ) \circ ( \textbf{p} v_0)
\nonumber \\
&& + ( \textbf{g} / v_0 ) \circ ( p^0 \textbf{e}_0 ) - \textbf{b} \times ( p^q \textbf{e}_q )
\nonumber \\
&& + ( \textbf{E} / v_0 ) \circ ( P^0 \textbf{E}_0 ) - \textbf{B} \times ( P^q \textbf{E}_q )     ~,
\end{eqnarray}
where $- \nabla \circ (p^0 v_0 \textbf{e}_0)$ is the energy gradient. $- ( \textbf{e}_0 \triangledown_0 ) \circ ( \textbf{p} v_0)$ is the inertial force. $ \{ ( \textbf{g} / v_0 ) \circ ( p^0 \textbf{e}_0 )  - \textbf{b} \times ( p^q \textbf{e}_q ) \}$ is the gravitational force. $ \{ ( \textbf{E} / v_0 ) \circ ( P^0 \textbf{E}_0 ) - \textbf{B} \times ( P^q \textbf{E}_q ) \}$ is the electromagnetic force. The product, $\textbf{E}_j \circ \textbf{E}_k$, belongs to the complex quaternion curved space, according to the multiplication of octonion.

When the octonion field potential is chosen as the first-rank tensor in the complex octonion curved space, the octonion linear momentum, angular momentum, and torque will be involved in the some spatial parameters of the complex octonion curved space (Table 3). It means that the curved space has an influence on the octonion force, including the force, the mass continuity equation, and the current continuity equation.

In general, the force $\textbf{N}$ consists of two components, in the complex octonion curved space. One is the force term $\textbf{N}_{flat}$ in the flat space, while the other is the additional force term $\textbf{N}_{curved}$ , caused by the curved space. Under most circumstances, the bending degree of curved space is quite tiny. As a result, the additional force term $\textbf{N}_{curved}$ is weak enough, while the force term $\textbf{N}_{flat}$ dominates the force $\textbf{N}$. Additionally, under some extreme conditions, the additional force term $\textbf{N}_{curved}$ may be strong enough, and even account for the main part.

\section{Conclusions}

In the complex quaternion curved space, from the definitions of quaternion metric, orthogonality, parallel translation, and covariant derivative, it is able to deduce the field potential, field strength, field source, linear momentum, angular momentum, torque, and force and so on in the gravitational field. The force includes the inertial force, gravitational force, and additional force term caused by the curved space. The connection coefficient and curvature of the curved space will act on the additional force term. When the gravitational potential is chosen as the first-rank tensor, the connection coefficient impacts the gravitational strength, while the curvature affects the gravitational strength and source.

In the complex octonion curved space, from the definitions of octonion metric, orthogonality, parallel translation, and covariant derivative, it is able to infer the octonion field potential, field strength, field source, linear momentum, angular momentum, torque, and force and so on in the gravitational and electromagnetic fields. The force consists of the inertial force, gravitational force, electromagnetic force, energy gradient, and additional force term caused by the complex octonion curved space. The connection coefficient and curvature of the complex octonion curved space may impact the additional force term. The study reveals that one may appraise the deviation amplitude of the complex octonion curved space departure from its flat space, by means of the measurements of the field potential, field strength, and force and so forth.

Specially, in the complex $S$-quaternion curved space for the electromagnetic field, it is capable of deducing the electromagnetic potential, strength, and source, and the additional force term caused by the curved space. The connection coefficient and curvature of the complex $S$-quaternion curved space have an influence on the additional force term. If the electromagnetic potential is the first-rank tensor, the connection coefficient of the complex $S$-quaternion curved space impacts the electromagnetic strength, while the curvature of the complex $S$-quaternion curved space affects the electromagnetic strength and source. Moreover the connection coefficient and curvature of the complex quaternion curved space will act on the electromagnetic strength and source as well.

It should be noted that the paper discussed only some simple cases about the influences of the complex octonion curved space on the field potential, field strength, and force and so forth. However it clearly states that the connection coefficient, curvature and other spatial parameters of the curved space exert an influence on the physical properties of gravitational and electromagnetic fields. In the following study, it is going to explore the impact of the force, in the strong gravitational and electromagnetic fields, on the movement statues of one charged objective in the complex octonion curved space. Moreover, it may intend to describe the gravitational and electromagnetic theories with any kind of frame, and then transform the field theories into that in the orthogonal affine frame by means of the appropriate transformation.

\section*{Acknowledgements}
The author is indebted to the anonymous referee for their valuable and constructive comments on the previous manuscript. This project was supported partially by the National Natural Science Foundation of China under grant number 60677039.

\appendix

\section{Basic postulates}

From the following basic postulates \cite{weng4}, the algebra of quaternions/octonions can be introduced into the field theory, describing some physical properties of electromagnetic and gravitational fields.

The first postulate: individual space. The space extended from the electromagnetic field is independent to one from the gravitational field. According to the viewpoint of Descartes \emph{et al.}, the fundamental field (electromagnetic or gravitational field) is an irreducible element of physical description, while the space is only the extension of the fundamental field and does not claim existence on its own. Further each fundamental field extends its individual space. These spaces are quite similar but independent to each other.

The second postulate: quaternion space. Each space extended respectively from the fundamental field can be chosen as the quaternion space. According to the viewpoint of Hamilton \emph{et al.}, the space in the physics can be chosen as the quaternion space. Two spaces, extended from the electromagnetic and gravitational fields respectively, can be considered as the quaternion spaces perpendicular to each other, so that they can combine together to become an octonion space. Consequently the octonion space can be applied to describe the physical properties of electromagnetic and gravitational fields.

The third postulate: dimensional homogeneity. It is necessary to introduce a few coefficients to maintain the dimensional homogeneity for all of physical quantities in each equation. In order to maintain the dimensional homogeneity, the physical quantity of electromagnetic field should be multiplied by a coefficient, when there are simultaneously physical quantities of electromagnetic and gravitational fields in an octonion equation. The coefficient can be determined by comparing with the classical theories relevant to the electromagnetic and gravitational fields.

\section{Connection coefficient}

The complex quaternion and octonion both are the first-rank tensors, according to the definition of tensor. Consequently it is possible to apply the tensor analysis to study various properties of complex octonion curved space, and the influence of curved space on physical quantities of gravitational and electromagnetic fields.

The complex octonion manifold is an indefinite geometric object, and its geometric structure is quite meager. A sole geometric property is that it is allowed to establish a coordinate system in the manifold, and transform the coordinate system with a continuous-differential function.

In the complex octonion manifold, it is possible to define the connection among different tensors in contiguous points, enabling the complex octonion manifold to become an affine connection space. In the paper, the torsion tensor of affine connection space is chosen to be zero. It means that the connection coefficient, $\Gamma^\beta_{\alpha \gamma}$ , is symmetric about the subscripts, $\alpha$ and $\gamma$ . That is, $\Gamma^\beta_{\alpha \gamma}$ = $\Gamma^\beta_{\gamma \alpha}$ . Apparently the coordinate value, connection coefficient, and octonion metric are all scalar, according to Eqs.(1), (2), (8), and (9), in the orthogonal affine frame system of the complex octonion curved space.

After introducing the connection, $\Gamma^\beta_{\alpha \gamma}$ , we can define the parallel translation now. The parallel translation is one geometric property irrespective of the choice of coordinate system. Let us transfer a complex octonion quantity, $\mathbb{Y} = Y^\beta \textbf{e}_\beta$ , from a point $M_1$ to the next point $M_2$ , requiring the quantity $\mathbb{Y}$ to remain unchanged. It means that the differential of quantity $\mathbb{Y}$ equals to zero. Therefore,
\begin{equation}
0 = ( d Y^\beta ) \textbf{e}_\beta  +  Y^\beta ( d \textbf{e}_\beta )  ~,
\end{equation}
or
\begin{equation}
0 = ( d Y^\beta ) \textbf{e}_\beta  +  Y^\beta ( \partial^2 \mathbb{H} / \partial u^\beta \partial u^\gamma ) d u^\gamma    ~,
\end{equation}
while the following term in the above can be extended into,
\begin{equation}
\partial^2 \mathbb{H} / \partial u^\beta \partial u^\gamma = \Gamma^\alpha_{\beta \gamma} \textbf{e}_\alpha  ~ .
\end{equation}

Substituting the above in Eq.(B.2) yields the condition of parallel translation for the complex octonion quantity $\mathbb{Y}$ ,
\begin{equation}
d Y^\beta = - \Gamma^\beta_{\alpha \gamma} Y^\alpha d u^\gamma   ~,
\end{equation}
and then, according to the property which the differential of a scalar equals to zero, $d (Y^\beta Y_\beta) = 0$ , one can obtain the equation of covariant component,
\begin{equation}
d Y_\beta = \Gamma^\alpha_{\beta \gamma} Y_\alpha d u^\gamma   ~.
\end{equation}

Further we can define the complex octonion metric in the affine connection space, achieving the complex octonion curved space, in possession of the parallel translation and metric. There is a certain relationship between the connection with metric. And that the definition of metric must meet the demand of this relationship in the affine connection space.

Multiplying Eq.(B.3) by the basis vector, $\textbf{e}_\lambda^* = \partial \mathbb{H}^* / \partial u^\lambda $ , from the left side can yield,
\begin{equation}
(\partial \mathbb{H}^* / \partial u^\lambda) \circ (\partial^2 \mathbb{H} / \partial u^\beta \partial u^\gamma) = g_{\overline{\lambda} \alpha} \Gamma^\alpha_{\beta \gamma}   ~ ,
\end{equation}
while multiplying the octonion conjugate of Eq.(B.3) by the basis vector, $\textbf{e}_\lambda$ , from the right side produces,
\begin{equation}
(\partial^2 \mathbb{H}^* / \partial u^\beta \partial u^\gamma) \circ (\partial \mathbb{H} / \partial u^\lambda) = \overline{\Gamma^\alpha_{\beta \gamma}} g_{\overline{{\alpha}} \lambda}  ~ ,
\end{equation}
where $\partial^2 \mathbb{H}^* / \partial u^\beta \partial u^\gamma = \overline{\Gamma^\alpha_{\beta \gamma}} \textbf{e}_\alpha^* $ . According to the octonion-Hermitian of metric, there is, $g_{\overline{\lambda} \alpha} = g_{\overline{\alpha} \lambda}$ , with $ [ ( g_{\overline{\lambda} \alpha} )^* ]^T = ( g_{\lambda \overline{\alpha}} )^T = g_{\overline{\alpha} \lambda} $ . The superscript $T$ denotes the transpose of matrix. $\Gamma^\alpha_{\beta \gamma}$ and $\overline{\Gamma^\alpha_{\beta \gamma}}$  both are coefficients.

Taking partial derivative of the metric tensor, $g_{\overline{\lambda} \gamma} = \textbf{e}_\lambda^* \circ \textbf{e}_\gamma$ , with respect to the coordinate $u^\beta$ from the left side achieves,
\begin{eqnarray}
&&(\partial \mathbb{H}^* / \partial u^\lambda ) \circ (\partial^2 \mathbb{H} / \partial u^\beta \partial u^\gamma)
\nonumber
\\
&& ~~~~~~~~+ (\partial^2 \mathbb{H}^* / \partial u^\beta \partial u^\lambda ) \circ (\partial \mathbb{H} / \partial u^\gamma)
= \partial g_{\overline{{\lambda}} \gamma} / \partial u^\beta   ~ ,
\end{eqnarray}
or
\begin{equation}
g_{\overline{\lambda} \alpha} \Gamma^\alpha_{\beta \gamma} + \overline{\Gamma^\alpha_{\beta \lambda}} g_{\overline{\alpha} \gamma}  = \partial g_{\overline{{\lambda}} \gamma} / \partial u^\beta   ~ .
\end{equation}

Similarly there are,
\begin{eqnarray}
&& g_{\overline{\beta} \alpha} \Gamma^\alpha_{\gamma \lambda} + \overline{\Gamma^\alpha_{\gamma \beta}} g_{\overline{\alpha} \lambda}  = \partial g_{\overline{{\beta}} \lambda} / \partial u^\gamma   ~ ,
\\
&& g_{\overline{\gamma} \alpha} \Gamma^\alpha_{\lambda \beta} + \overline{\Gamma^\alpha_{\lambda \gamma}} g_{\overline{\alpha} \beta}  = \partial g_{\overline{{\gamma}} \beta} / \partial u^\lambda   ~ .
\end{eqnarray}

From the last three equations in the above, we have,
\begin{eqnarray}
&& [ ( g_{\overline{\lambda} \alpha} \Gamma^\alpha_{\beta \gamma} )^T ]^* +  [ ( \overline{\Gamma^\alpha_{\gamma \beta}} g_{\overline{\alpha} \lambda} )^T ]^*
\nonumber
\\
&& ~~~~~~~~ = ( \partial g_{\overline{\gamma} \lambda} / \partial u^\beta + \partial g_{\overline{\lambda} \beta} / \partial u^\gamma - \partial g_{\overline{\gamma} \beta} / \partial u^\lambda )  ~ .
\end{eqnarray}

Because $ [ ( g_{\overline{\lambda} \alpha} \Gamma^\alpha_{\beta \gamma} )^T ]^* =  \overline{\Gamma^\alpha_{\gamma \beta}} g_{\overline{\alpha} \lambda} $ , $ [ ( \overline{\Gamma^\alpha_{\gamma \beta}} g_{\overline{\alpha} \lambda} )^T ]^* = g_{\overline{\lambda} \alpha} \Gamma^\alpha_{\beta \gamma} $ , there is,
\begin{eqnarray}
g_{\overline{\lambda} \alpha} \Gamma^\alpha_{\beta \gamma} + [ ( g_{\overline{\lambda} \alpha} \Gamma^\alpha_{\beta \gamma} )^T ]^*
= \overline{\Gamma^\alpha_{\gamma \beta}} g_{\overline{\alpha} \lambda} + [ ( \overline{\Gamma^\alpha_{\gamma \beta}} g_{\overline{\alpha} \lambda} )^T ]^* ~,
\end{eqnarray}
it means that, $ g_{\overline{\lambda} \alpha} \Gamma^\alpha_{\beta \gamma} =  \overline{\Gamma^\alpha_{\gamma \beta}} g_{\overline{\alpha} \lambda} $ , therefore Eq.(B.12) can be rewritten as,
\begin{eqnarray}
\Gamma_{\overline{\lambda}, \beta \gamma} = (1/2) ( \partial g_{\overline{\gamma} \lambda} / \partial u^\beta + \partial g_{\overline{\lambda} \beta} / \partial u^\gamma - \partial g_{\overline{\gamma} \beta} / \partial u^\lambda )  ~ ,
\end{eqnarray}
where $ \Gamma_{\overline{\lambda}, \beta \gamma} = g_{\overline{\lambda} \alpha} \Gamma^\alpha_{\beta \gamma} $ . $\Gamma^\alpha_{\beta \gamma} = g^{\alpha \overline{\lambda}} \Gamma_{\overline{\lambda}, \beta \gamma}$ , with $  g^{ \alpha \overline{\lambda}} g_{ \overline{\lambda} \beta} = \delta^\alpha_\beta $ . $\Gamma^\alpha_{\gamma \beta} = \Gamma^\alpha_{\beta \gamma} $ .

Multiplying Eq.(B.3) by the basis vector, $\textbf{e}_\lambda^* = \partial \mathbb{H}^* / \partial u^\lambda $ , from the left side can yield,
\begin{equation}
(\partial \mathbb{H}^* / \partial u^\lambda) \circ (\partial^2 \mathbb{H} / \partial \overline{u^\beta} \partial u^\gamma) = g_{\overline{\lambda} \alpha} \Gamma^\alpha_{\overline{\beta} \gamma}   ~ ,
\end{equation}
while multiplying the octonion conjugate of Eq.(B.3) by the basis vector, $\textbf{e}_\lambda$ , from the right side produces,
\begin{equation}
(\partial^2 \mathbb{H}^* / \partial \overline{u^\beta} \partial u^\gamma) \circ (\partial \mathbb{H} / \partial u^\lambda) = \overline{\Gamma^\alpha_{\overline{\beta} \gamma}} g_{\overline{{\alpha}} \lambda}  ~ ,
\end{equation}
where $\partial^2 \mathbb{H} / \partial \overline{u^\beta} \partial u^\gamma = \Gamma^\alpha_{\overline{\beta} \gamma} \textbf{e}_\alpha $ , and $\partial^2 \mathbb{H}^* / \partial \overline{u^\beta} \partial u^\gamma = \overline{\Gamma^\alpha_{\overline{\beta} \gamma}} \textbf{e}_\alpha^* $ . $\Gamma^\alpha_{\overline{\beta} \gamma}$ and $\overline{\Gamma^\alpha_{\overline{\beta} \gamma}} $ both are coefficients.

Taking partial derivative of the metric tensor, $g_{\overline{\lambda} \gamma} = \textbf{e}_\lambda^* \circ \textbf{e}_\gamma$ , with respect to the coordinate $\overline{u^\beta}$ from the left side achieves,
\begin{eqnarray}
&&(\partial \mathbb{H}^* / \partial u^\lambda ) \circ (\partial^2 \mathbb{H} / \partial \overline{u^\beta} \partial u^\gamma)
\nonumber
\\
&& ~~~~~~~~+ (\partial^2 \mathbb{H}^* / \partial \overline{u^\beta} \partial u^\lambda ) \circ (\partial \mathbb{H} / \partial u^\gamma)
= \partial g_{\overline{{\lambda}} \gamma} / \partial \overline{u^\beta}   ~ ,
\end{eqnarray}
or
\begin{equation}
g_{\overline{\lambda} \alpha} \Gamma^\alpha_{\overline{\beta} \gamma} + \overline{\Gamma^\alpha_{\overline{\beta} \lambda}} g_{\overline{\alpha} \gamma}  = \partial g_{\overline{{\lambda}} \gamma} / \partial \overline{u^\beta}   ~ .
\end{equation}

Similarly there are,
\begin{eqnarray}
&& g_{\overline{\beta} \alpha} \Gamma^\alpha_{\overline{\gamma} \lambda} + \overline{\Gamma^\alpha_{\overline{\gamma} \beta}} g_{\overline{\alpha} \lambda}  = \partial g_{\overline{{\beta}} \lambda} / \partial \overline{u^\gamma}   ~ ,
\\
&& g_{\overline{\gamma} \alpha} \Gamma^\alpha_{\overline{\lambda} \beta} + \overline{\Gamma^\alpha_{\overline{\lambda} \gamma}} g_{\overline{\alpha} \beta}  = \partial g_{\overline{{\gamma}} \beta} / \partial \overline{u^\lambda}   ~ .
\end{eqnarray}

From the last three equations in the above, we have,
\begin{eqnarray}
&& [ ( g_{\overline{\lambda} \alpha} \Gamma^\alpha_{\overline{\beta} \gamma} )^T ]^* +  [ ( \overline{\Gamma^\alpha_{\overline{\gamma} \beta}} g_{\overline{\alpha} \lambda} )^T ]^*
\nonumber
\\
&& ~~~~~~~~ = ( \partial g_{\overline{\gamma} \lambda} / \partial \overline{u^\beta} + \partial g_{\overline{\lambda} \beta} / \partial \overline{u^\gamma} - \partial g_{\overline{\gamma} \beta} / \partial \overline{u^\lambda} )  ~ .
\end{eqnarray}

Because $ [ ( g_{\overline{\lambda} \alpha} \Gamma^\alpha_{\overline{\beta} \gamma} )^T ]^* =  \overline{\Gamma^\alpha_{\overline{\gamma} \beta}} g_{\overline{\alpha} \lambda} $ , $ [ ( \overline{\Gamma^\alpha_{\overline{\gamma} \beta}} g_{\overline{\alpha} \lambda} )^T ]^* = g_{\overline{\lambda} \alpha} \Gamma^\alpha_{\overline{\beta} \gamma} $ , there is,
\begin{eqnarray}
g_{\overline{\lambda} \alpha} \Gamma^\alpha_{\overline{\beta} \gamma} + [ ( g_{\overline{\lambda} \alpha} \Gamma^\alpha_{\overline{\beta} \gamma} )^T ]^*
= \overline{\Gamma^\alpha_{\overline{\gamma} \beta}} g_{\overline{\alpha} \lambda} + [ ( \overline{\Gamma^\alpha_{\overline{\gamma} \beta}} g_{\overline{\alpha} \lambda} )^T ]^* ~,
\end{eqnarray}
it means that, $ g_{\overline{\lambda} \alpha} \Gamma^\alpha_{\overline{\beta} \gamma} =  \overline{\Gamma^\alpha_{\overline{\gamma} \beta}} g_{\overline{\alpha} \lambda} $ , therefore Eq.(B.21) can be rewritten as,
\begin{eqnarray}
\Gamma_{\overline{\lambda}, \overline{\beta} \gamma} = (1/2) ( \partial g_{\overline{\gamma} \lambda} / \partial \overline{u^\beta} + \partial g_{\overline{\lambda} \beta} / \partial \overline{u^\gamma} - \partial g_{\overline{\gamma} \beta} / \partial \overline{u^\lambda} )  ~ ,
\end{eqnarray}
where $ \Gamma_{\overline{\lambda}, \overline{\beta} \gamma} = g_{\overline{\lambda} \alpha} \Gamma^\alpha_{\overline{\beta} \gamma} $ . $\Gamma^\alpha_{\overline{\beta} \gamma} = g^{\alpha \overline{\lambda}} \Gamma_{\overline{\lambda}, \overline{\beta} \gamma}$ , with $ [ ( \Gamma^\alpha_{\overline{\beta} \gamma} )^* ]^T  = \Gamma^\alpha_{\overline{\gamma} \beta} $ .
$\Gamma^\alpha_{\gamma \overline{\beta}} = \Gamma^\alpha_{\overline{\beta} \gamma} $ .

In the complex octonion curved space, the covariant derivative can be defined as Eq.(9) and
\begin{equation}
\triangledown_{\overline{\gamma}} Y^\alpha = \partial ( \delta^\alpha_\beta Y^\beta ) / \partial \overline{u^\gamma} + \Gamma^\alpha_{\beta \overline{\gamma}} Y^\beta   ~ .
\end{equation}

Furthermore, for a mixed tensor, $Y_\nu^\gamma$ , with contravariant rank 1 and covariant rank 1, the covariant derivative can be written as,
\begin{eqnarray}
&& \triangledown_\beta Y_\nu^\gamma = \partial Y_\nu^\gamma / \partial u^\beta - \Gamma^\lambda_{\beta \nu} Y_\lambda^\gamma + \Gamma^\gamma_{\beta \lambda} Y_\nu^\lambda  ~ ,
\\
&& \triangledown_{\overline{\alpha}} Y_\nu^\gamma = \partial Y_\nu^\gamma / \partial \overline{u^\alpha} - \Gamma^\lambda_{\overline{\alpha} \nu} Y_\lambda^\gamma + \Gamma^\gamma_{\overline{\alpha} \lambda} Y_\nu^\lambda  ~ .
\end{eqnarray}

Therefore, one can find that,
\begin{eqnarray}
\triangledown_{\overline{\alpha}} ( \triangledown_\beta Y^\gamma ) - \triangledown_\beta ( \triangledown_{\overline{\alpha}} Y^\gamma ) = R_{\beta \overline{\alpha} \nu}^{~~~~\gamma} Y^\nu + T^\lambda_{\beta \overline{\alpha}} (\triangledown_\lambda Y^\gamma)  ~,
\end{eqnarray}
with
\begin{eqnarray}
R_{\beta \overline{\alpha} \nu}^{~~~~\gamma} = \partial \Gamma^\gamma_{\nu \beta} / \partial \overline{u^\alpha} - \partial \Gamma^\gamma_{\nu \overline{\alpha}} / \partial u^\beta  +  \Gamma^\gamma_{\lambda \overline{\alpha}} \Gamma^\lambda_{\nu \beta} - \Gamma^\gamma_{\lambda \beta} \Gamma^\lambda_{\nu \overline{\alpha}}  ~, ~~
\end{eqnarray}
where the curvature tensor is $R_{\beta \overline{\alpha} \nu}^{~~~~\gamma}$ , while $R_{\beta \overline{\alpha} \nu \overline{\lambda}}  = R_{\beta \overline{\alpha} \nu}^{~~~~\gamma} g_{\gamma \overline{\lambda}}$ . And the torsion tensor is, $T^\lambda_{\beta \overline{\alpha}} = \Gamma^\lambda_{\overline{\alpha} \beta} - \Gamma^\lambda_{\beta \overline{\alpha}} $ . If $T^\lambda_{\beta \overline{\alpha}} = 0$, there is, $\Gamma^\lambda_{\overline{\alpha} \beta} = \Gamma^\lambda_{\beta \overline{\alpha}} $ .

And now we can discuss some physical properties of gravitational and electromagnetic fields in the complex octonion space, by means of the covariant derivative and covariant differential in the above.

In case the complex $S$-quaternion radius vector, $k_{eg} \mathbb{H}_e$ , is able to be neglected approximately, the complex octonion radius vector, $\mathbb{H}$ , will be degenerated into the complex quaternion radius vector, $\mathbb{H}_g$ . Consequently the complex octonion curved space will be reduced into the complex quaternion curved space, while the previous discussions in the complex octonion curved space will be reduced into that in the complex quaternion curved space.

\section{Gravitational field equations}

In the complex quaternion curved space for the gravitational field, the gravitational potential, $\mathbb{A}_g ( i a^0 , a^1 , a^2 , a^3 )$, is defined as Eq.(3), or
\begin{equation}
\mathbb{A}_g = i \lozenge^\star \circ \mathbb{X}_g   ~ ,
\end{equation}
where $\mathbb{A}_g = i \lozenge^\star \circledcirc \mathbb{X}_g + i \lozenge^\star \circledast \mathbb{X}_g$ . Choose one appropriate gauge condition to insure that, the coordinate of the scalar part, $i \lozenge^\star \circledcirc \mathbb{X}_g$ , contains only the imaginary part, while the coordinate of the vector part, $i \lozenge^\star \circledast \mathbb{X}_g$ , covers merely the real part.

The gravitational strength, $\mathbb{F}_g ( i f^0 , f^1 , f^2 , f^3 )$, is Eq.(4), or
\begin{equation}
\mathbb{F}_g = \lozenge\circ \mathbb{A}_g  ~,
\end{equation}
where $\mathbb{F}_g = \lozenge \circledcirc \mathbb{A}_g + \lozenge \circledast \mathbb{A}_g$ . The coordinate of the vector part, $\lozenge \circledast \mathbb{A}_g$ , is one complex number. Choose one appropriate gauge condition to insure that, the coordinate of the scalar part, $\lozenge \circledcirc \mathbb{A}_g$ , equals to zero.

The gravitational source, $\mathbb{S}_g ( i s^0 , s^1 , s^2 , s^3 )$, is Eq.(6), or
\begin{equation}
- \mu_g \mathbb{S}_g = \lozenge^* \circ \mathbb{F}_g  ~  ,
\end{equation}
where $- \mu_g \mathbb{S}_g = \lozenge^* \circledcirc \mathbb{F}_g + \lozenge^* \circledast \mathbb{F}_g $ .

In the above, the scalar part of $\mathbb{S}_g$ is, $- \lozenge^* \circledcirc \mathbb{F}_g / \mu_g = i s^0 \textbf{e}_0$ , and its coordinate is one complex number. Further the scalar part can be rewritten as,
\begin{equation}
\nabla^* \cdot ( i \textbf{g} / v_0 + \textbf{b} ) = - i \mu_g s^0 \textbf{e}_0  ~ ,
\end{equation}
and expanding the above yields two component equations,
\begin{eqnarray}
& \nabla \cdot \textbf{b} = 0  ~ ,
\\
& \nabla^* \cdot ( \textbf{g} / v_0 ) = - \mu_g s^0 \textbf{e}_0  ~ .
\end{eqnarray}

Meanwhile the vector part of $\mathbb{S}_g$ is, $- \lozenge^* \circledast \mathbb{F}_g / \mu_g = s^r \textbf{e}_r$ , and its coordinate is one complex number. And that the scalar part can be rewritten as,
\begin{equation}
i (\textbf{e}_0 \triangledown_0) \circ ( i \textbf{g} / v_0 + \textbf{b} ) + \nabla^* \times ( i \textbf{g} / v_0 + \textbf{b} ) = - \mu_g s^r \textbf{e}_r  ~ ,
\end{equation}
also expanding the above yields two component equations,
\begin{eqnarray}
& (\textbf{e}_0 \triangledown_0) \circ \textbf{b} + \nabla^* \times ( \textbf{g} / v_0 ) = 0   ~ ,
\\
& - (\textbf{e}_0 \triangledown_0) \circ ( \textbf{g} / v_0 ) + \nabla^* \times \textbf{b} = - \mu_g s^r \textbf{e}_r   ~ .
\end{eqnarray}

These four component equations, Eqs.(C.5), (C.6), (C.8), and (C.9), are the gravitational field equations in the complex quaternion curved space. When $\textbf{a} = 0$ and $\textbf{b} = 0$, one of equations will be simplified into the Newton's law of universal gravitation in the classical gravitational theory. The influence of the linear momentum, $(s^r \textbf{e}_r )$, can be ignored in general, for the gravitational constant $\mu_g$ is weak in the extreme.

In the complex quaternion curved space, the definition of gravitational source can be written as,
\begin{equation}
- \mu_g \mathbb{S}_g = \lozenge^* \circ ( \lozenge \circ \mathbb{A}_g) = \lozenge^2  \mathbb{A}_g   ~ ,
\end{equation}
and it is able to be separated into two component equations,
\begin{eqnarray}
& - \mu_g s^0 \textbf{e}_0 = \lozenge^2  a^0 \textbf{e}_0  ~  ,
\\
& - \mu_g s^r \textbf{e}_r = \lozenge^2  a^q \textbf{e}_q  ~ ,
\end{eqnarray}
where $\lozenge^2 \mathbb{A}_g = \lozenge^* \circ ( \lozenge \circ \mathbb{A}_g) $ .

These two component equations are the d'Alembert equations of gravitational field in the complex quaternion curved space. It means that the d'Alembert equations and the gravitational field equations are involved in the spatial parameters (connection coefficient and curvature and so on) of the complex quaternion curved space, when the gravitational potential is chosen as the first-rank tensor.

In the complex quaternion curved space, the definition of the gravitational strength can be written as,
\begin{equation}
\mathbb{F}_g = \lozenge \circ ( i \lozenge^\star \circ \mathbb{X}_g)   ~ ,
\end{equation}
further the above may be rewritten as,
\begin{equation}
\mathbb{F}_g = i \{ \triangledown_0 \triangledown_0 + \nabla \circ \nabla  + i ( \triangledown_0 \nabla - \nabla \triangledown_0 ) \} \circ \mathbb{X}_g   ~ .
\end{equation}

Similarly the gravitational strength is involved in the spatial parameters of complex quaternion curved space, when the integrating function of gravitational potential is chosen as the first-rank tensor.

\section{Electromagnetic field equations}

In the complex $S$-quaternion curved space, the electromagnetic potential, $\mathbb{A}_e ( i A^0 , A^1 , A^2 , A^3 )$, is
\begin{equation}
\mathbb{A}_e = i \lozenge^\star \circ \mathbb{X}_e   ~ ,
\end{equation}
where $\mathbb{A}_e = i \lozenge^\star \circledcirc \mathbb{X}_e + i \lozenge^\star \circledast \mathbb{X}_e $ . Choose one proper gauge condition to make sure that, the coordinate of the `scalar' part, $i \lozenge^\star \circledcirc \mathbb{X}_e$ , contains only the imaginary part, while the coordinate of the vector part, $i \lozenge^\star \circledast \mathbb{X}_e$ , covers merely the real part.

The electromagnetic strength, $\mathbb{F}_e ( i F^0 , F^1 , F^2 , F^3 )$ , is
\begin{equation}
\mathbb{F}_e = \lozenge \circ \mathbb{A}_e   ~ ,
\end{equation}
where $\mathbb{F}_e =  \lozenge \circledcirc \mathbb{A}_e + \lozenge \circledast \mathbb{A}_e $ . The coordinate of the vector part, $\lozenge \circledast \mathbb{A}_e$ , is one complex number. Choose one appropriate gauge condition to insure that, the coordinate of the `scalar' part, $\lozenge \circledcirc \mathbb{A}_e$ , equals to zero.

The electromagnetic source, $\mathbb{S}_e ( i S^0 , S^1 , S^2 , S^3 )$ , is
\begin{equation}
- \mu_e \mathbb{S}_e = \lozenge^* \circ \mathbb{F}_e   ~ ,
\end{equation}
where $- \mu_e \mathbb{S}_e =  \lozenge^* \circledcirc \mathbb{F}_e + \lozenge^* \circledast \mathbb{F}_e $ .

In the above, the `scalar' part of $\mathbb{S}_e$ is, $- \lozenge^* \circledcirc \mathbb{F}_e  / \mu_e = i S^0 \textbf{E}_0$, and its coordinate is one complex number. Further the `scalar' part can be rewritten as,
\begin{equation}
\nabla^* \cdot ( i \textbf{E} / v_0 + \textbf{B} ) = - i \mu_e S^0 \textbf{E}_0  ~ ,
\end{equation}
and expanding the above yields two component equations,
\begin{eqnarray}
& \nabla \cdot \textbf{B} = 0  ~ ,
\\
& \nabla^* \cdot ( \textbf{E} / v_0 ) = - \mu_e S^0 \textbf{E}_0  ~  .
\end{eqnarray}

Meanwhile the vector part of $\mathbb{S}_e$ is, $- \lozenge^* \circledast \mathbb{F}_e / \mu_e = S^r \textbf{E}_r$ , and its coordinate is one complex number. And that the `scalar' part can be rewritten as,
\begin{equation}
i (\textbf{e}_0 \triangledown_0) \circ ( i \textbf{E} / v_0 + \textbf{B} ) + \nabla^* \times ( i \textbf{E} / v_0 + \textbf{B} ) = - \mu_e S^r \textbf{E}_r   ~ ,
\end{equation}
also expanding the above yields two component equations,
\begin{eqnarray}
& (\textbf{e}_0 \triangledown_0) \circ  \textbf{B} + \nabla^* \times ( \textbf{E} / v_0 ) = 0   ~ ,
\\
& - (\textbf{e}_0 \triangledown_0) \circ  ( \textbf{E} / v_0 ) + \nabla^* \times \textbf{B} = - \mu_e S^r \textbf{E}_r    ~ .
\end{eqnarray}

These four component equations, Eqs.(D.5), (D.6), (D.8), and (D.9), are the electromagnetic field equations in the complex $S$-quaternion curved space. When the curved space is degenerated into the flat space, the above will be reduced to the electromagnetic field equations in the complex $S$-quaternion flat space.

In the complex quaternion curved space, the definition of electromagnetic source can be written as,
\begin{equation}
- \mu_e \mathbb{S}_e = \lozenge^* \circ ( \lozenge \circ \mathbb{A}_e)  ~ ,
\end{equation}
and it is able to be separated into two component equations,
\begin{eqnarray}
& - \mu_e S^0 \textbf{E}_0 = \lozenge^* \circ ( \lozenge \circ  A^0 \textbf{E}_0 )  ~ ,
\\
& - \mu_e S^r \textbf{E}_r = \lozenge^* \circ ( \lozenge \circ  A^q \textbf{E}_q )  ~ .
\end{eqnarray}

These two component equations are the d'Alembert equations of the electromagnetic field in the complex $S$-quaternion curved space. It means that the d'Alembert equations and the electromagnetic field equations are associated with the spatial parameters of complex octonion (but not limited to the complex $S$-quaternion) curved space, when the electromagnetic potential is chosen as the first-rank tensor.

In the complex $S$-quaternion curved space, the definition of electromagnetic strength can be written as,
\begin{eqnarray}
\mathbb{F}_e = \lozenge \circ (i \lozenge^\star \circ \mathbb{X}_e)   ~ ,
\end{eqnarray}
further it may be rewritten as,
\begin{equation}
\mathbb{F}_e = i \{ \triangledown_0 \triangledown_0 + \nabla \circ \nabla \} \circ \mathbb{X}_e - \{ \triangledown_0 ( \nabla \circ \mathbb{X}_e ) - \nabla \circ ( \triangledown_0 \mathbb{X}_e ) \} ~ .
\end{equation}

Similarly the electromagnetic strength is involved in the spatial parameters of complex octonion curved space, when the integrating function of electromagnetic potential is chosen as the first-rank tensor.


\begin{thebibliography}{0}


\bibitem{morita}
      K. Morita,
      Quaternions, Lorentz group and the Dirac theory,
      {\it Prog. Theor. Phys.\/},
      \textbf{117} (3) (2007) 501--532.

\bibitem{edmonds}
      J. D. Edmonds,
      Quaternion wave equations in curved space-time,
      {\it Int. J. Theor. Phys.\/},
      \textbf{10} (2) (1974) 115--122.

\bibitem{majernik}
      V. Majernik,
      Quaternionic formulation of the classical fields,
      {\it Adv. Appl. Clifford Al.\/},
      \textbf{9} (1) (1999) 119--130.

\bibitem{rawat}
      A. S. Rawat and O. P. S. Negi,
      Quaternion gravi-electromagnetism,
      {\it Int. J. Theor. Phys.\/},
      \textbf{51} (3) (2012) 738--745.

\bibitem{gogberashvili}
      M. Gogberashvili,
      Octonionic electrodynamics,
      {\it J. Phys. A\/},
      \textbf{39} (22) (2006) 7099--7104.

\bibitem{demir}
      S. Demir, M. Tanisli, and T. Tolan,
      Octonionic gravitational field equations,
      {\it Int. J. Mod. Phys. A\/},
      \textbf{28} (21) (2013) 1350112, 3 pages.

\bibitem{mironov}
      V. L. Mironov and S. V. Mironov,
      Octonic representation of electromagnetic field equations,
      {\it J. Math. Phys.\/},
      \textbf{50} (1) (2009) 012901, 10 pages.

\bibitem{negi}
      B. C. Chanyal, P. S. Bisht, and O. P. S. Negi,
      Generalized octonion electrodynamics,
      {\it Int. J. Theor. Phys.\/},
      \textbf{49} (6) (2010) 1333--1343.

\bibitem{weng1}
      Z.-H. Weng,
      Angular momentum and torque described with the complex octonion,
      {\it AIP Adv.\/},
      \textbf{4} (8) (2014) 087103, 16 pages.
      [Erratum: {\it ibid.} {\bf 5} (2015) 109901 ].

\bibitem{bishop}
      R. L. Bishop,
      There is more than one way to frame a curve,
      {\it Amer. Math. Monthly\/},
      \textbf{82} (3) (1975) 246--251.

\bibitem{morandi}
      P. J. Morandi, J. M. Perez-Izquierdo, and S. Pumplun,
      On the tensor product of composition algebras,
      {\it J. Algebra\/},
      \textbf{243} (1) (2001) 41--68.

\bibitem{marques}
      S. Marques-Bonham,
      The Dirac equation in a non-Riemannian manifold III: An analysis using the algebra of quaternions and octonions,
      {\it J. Math. Phys.\/},
      \textbf{32} (5) (1991) 1383--1394.

\bibitem{dundarer}
      A. R. Dundarer,
      Multi-instanton solutions in eight-dimensional curved space,
      {\it Mod. Phys. Lett. A\/},
      \textbf{6} (5) (2011) 409--415.

\bibitem{tsagas}
      C. G. Tsagas,
      Electromagnetic fields in curved spacetimes,
      {\it Classical Quant. Grav.\/},
      \textbf{22} (2) (2005) 393--407.

\bibitem{castro}
      C. Castro,
      On octonionic gravity, exceptional Jordan strings and nonassociative ternary Gauge field theories,
      {\it Int. J. Geom. Methods Mod. Phys.\/},
      \textbf{9} (2) (2012) 1250021, 23 pages.

\bibitem{demir2}
      S. Demir,
      Hyperbolic octonion formulation of gravitational field equations,
      {\it Int. J. Theor. Phys.\/},
      \textbf{52} (1) (2013) 105--116.

\bibitem{chanyal}
      B. C. Chanyal, P. S. Bisht, and O. P. S. Negi,
      Octonionic non-Abelian gauge theory,
      {\it Int. J. Theor. Phys.\/},
      \textbf{52} (10) (2013) 3522--3533.

\bibitem{kalauni}
      P. Kalauni and J. C. A. Barata,
      Reconstruction of symmetric Dirac-Maxwell equations using nonassociative algebra,
      {\it Int. J. Theor. Phys.\/},
      \textbf{12} (3) (2015) 1550029, 9 pages.

\bibitem{weng2}
      Z.-H. Weng,
      Field equations in the complex quaternion spaces,
      {\it Adv. Math. Phys.\/},
      \textbf{2014} (2014) 450262, 6 pages.

\bibitem{fischbach}
      Y. Bonder, E. Fischbach, H. Hernandez-Coronado, D. E. Krause, Z. Rohrbach, D. Sudarsky,
      Testing the equivalence principle with unstable particles,
      {\it Phys. Rev. D\/},
      \textbf{87} (12) (2013) 125021, 12 pages.

\bibitem{adelberger}
      E. G. Adelberger, C. W. Stubbs, B. R. Heckel, Y. Su, H. E. Swanson, G. L. Smith, J. H. Gundlach, and W. F. Rogers,
      Testing the equivalence principle in the field of the Earth: Particle physics at masses below 1 $\mu$ eV,
      {\it Phys. Rev. D\/},
      \textbf{42} (10) (1990) 3267--3292.

\bibitem{baessler}
      S. Baessler, B. R. Heckel, E. G. Adelberger, J. H. Gundlach, U. Schmidt, and H. E. Swanson,
      Improved test of the equivalence principle for gravitational self-Energy,
      {\it Phys. Rev. Lett.\/},
      \textbf{83} (18) (1999) 3585--3588.

\bibitem{turyshev}
      S. G. Turyshev, U. E. Israelsson, M. Shao, N. Yu, A. Kusenko, E. L. Wright, C. W. F. Everitt, M. Kasevich, J. A. Lipa, J. C. Mester, R. D. Reasenberg, R. L. Walsworth, N. Ashay, H. Gould, and H. J. Paik,
      Space-based research in fundamental physics and quantum technologies,
      {\it Int. J. Mod. Phys. D\/},
      \textbf{16} (12a) (2007) 1879--1925.

\bibitem{moffat}
      J. W. Moffat and G. T. Gillies,
      Satellite E\"{o}tv\"{o}s test of the weak equivalence principle for zero-point vacuum energy,
      {\it New J. Phys.\/},
      \textbf{4} (1) (2002) 92, 6 pages.

\bibitem{liu}
      M.-Y. Liu, Z.-H. Zhong, Y.-C. Han, X.-Y. Wang, Z.-S. Yang, and Y. Xie,
      Preliminary limits on deviation from the inverse-square law of gravity in the solar system: a power-law parameterization,
      {\it Res. Astron. Astrophys.\/},
      \textbf{14} (8) (2014) 1019--1028.

\bibitem{weng3}
      Z.-H. Weng,
      Dynamic of astrophysical jets in the complex octonion space,
      {\it Int. J. Mod. Phys. D\/},
      \textbf{24} (8) (2015) 1550072, 19 pages.

\bibitem{reasenberg}
      R. D. Reasenberg,
      A new class of equivalence principle test masses, with application to SR-POEM,
      {\it Classical Quant. Grav.\/},
      \textbf{31} (17) (2014) 175013, 15 pages.

\bibitem{weng4}
      Z.-H. Weng,
      Maxwell Equations of Electromagnetic and Gravitational Fields in the Curved Spaces,
      {\it PIERS Proceedings\/},
      Stockholm, Sweden, August 12-15, 2013, pp.984-988.


\end{thebibliography}
\end{document}